# PHYSICAL MECHANISMS INFLUENCING LIFE ORIGIN AND DEVELOPMENT. PHYSICAL-BIOCHEMICAL PARADIGM OF LIFE

YURI K. SHESTOPALOFF

The present view of biological phenomena is based on a biochemical paradigm that development of living organisms is defined by information stored in a molecular form as some genetic code. However, new facts and discoveries indicate that biological phenomena cannot be confined to a biochemical realm alone, but are also influenced by physical mechanisms. These mechanisms work at cellular, organ and whole organism spatial levels. They impose uniquely defined constraints on distribution of nutrients between biomass synthesis and maintenance of existing biomass, thus influencing the composition of biochemical reactions, their ssuccessive change and irreversibility during the organismal life cycle. Mathematically, such a growth mechanism is represented by a growth equation, which is mathematically an ordinary differential equation. Using this equation, we introduce growth models, show their adequacy to experimental data, and discover two types of division mechanisms, examining growth of unicellular organisms *Amoeba*, *S. pombe*, *E. coli*, *B. subtilis*, *Staphylococcus*. Also, on the basis of the growth equation, we find different metabolic characteristics of these organisms. For instance, it was shown that in logarithmic coordinates the values of their metabolic allometric exponents are located on a straight line. This fact has important implications with regard to evolutionary process of organisms within a food chain, considered as a single system. High adequateness of obtained results to experimental data, from different perspectives, as well as excellent compliance with previously proven more particular knowledge, and with general criteria for validation of scientific truths, proves validity of the introduced general growth mechanism and the growth equation. Taken together, the obtained results set solid grounds for introduction of a more comprehensive physical-biochemical paradigm of Life origin, development and evolution.

## 1. Introduction

### 1.1. *Origin and the present state of a biochemical paradigm of Life*

The main idea of this paper is introduction of certain *physical* mechanisms, acting at higher than molecular scale levels and participating in growth and division of living organisms. Such mechanisms work alongside and in tight cooperation with biochemical mechanisms. This view is in some contradiction with the present biochemical paradigm of life, according to which life origin and development is overwhelmingly due to workings of biochemical machinery. In order to understand the problem, we will begin from historical-philosophical excurse of how and on which basis this biochemical view originated, what assumptions of the original conception secured such a “biochemical monopoly”, and to what extent they were justified. One of the founders of a biochemical paradigm in biology is considered *Physicist* E. Schrödinger[1], whose views ware presented in public lectures delivered in 1943. He was probably the first who pronounced the words 'information' and 'code' relative to organismal molecular structures, although not exactly in such determinate meaning, which was later transformed into its present canonized form in textbooks. What he actually said, was "*In calling the structure of the chromosome fibres a code script we mean that the all penetrating mind, once conceived by Laplace, to which every causal connection lay immediately open, could tell from their structure whether the egg would develop, under suitable conditions, into a black cock or* *into a speckled hen, into a fly or a maize plant, a rhododendron, a beetle, a mouse or a woman. To which we may add, that the appearances of the egg cells are very often remarkably similar; and even when they are not, as in the case of the comparatively gigantic eggs of birds and reptiles, the difference is not been so much the relevant structures*". Compare the meaning of this quote to what is said in a canonical text Ref. 2: "*The cell-cycle control system is based on a connected series of biochemical switches, each of which initiates a specific cell-cycle event. This system of switches possesses many important engineering features that increase the accuracy and reliability of cell-cycle progression. First, the switches are generally binary (on/off) and launch events in a complete irreversible fashion*". The concept of a cell-cycle control presented in the second quote is rather loosely related to the original Schrödinger's thought, the distance is noticeable; the tonality of suggestion in the Schrödinger's quote disappeared, and the second statement is posited as an indisputable truth. On the other hand, to an unbiased observer, the described system of binary switches looks rather as an *execution* mechanism, but not as a *managing* system - sort of a household electrical system having an electrical panel with fuses, wires and switches to make the household system functional. However, these are the house occupants, which manage this system.

What is important to understand, both the quote from Ref. 2, and Schrödinger's thought, present *hypotheses*, which are both not invincible from the known facts and consistent logic. For instance, they do not provide an answer to the first question coming to mind - which mechanism triggers the very first switch (or the very first switches?). Furthermore, there are many parallel and intersecting branches of biochemical reactions. It is very difficult to imagine for such an *enormously* complex system to be governed synchronically in time and space by molecular events alone acting in a binary switch manner, without a general framework, standing above and uniting all these events into a single coherent process. And these are just two from a myriad of other unanswered principal questions.

The Schrödinger's problem was that he interpreted his *assumption* as a proven theory, while it was not. He says: "*For it is simply a fact of observation that the guiding principle in every cell is embodied in a single atomic association existing only one copy (or sometimes two) - and a fact of observation that it may results in producing events which are a paragon of orderliness. Whether we find it astonishing or whether we find it quite plausible that a small but highly organized group of atoms be capable of acting in this manner, the situation is unprecedented, it is unknown anywhere else except in living matter. The physicist and the chemist, investigating inanimate matter, have never witnessed phenomena which they had to interpret in this way*."

In fact, the aforementioned observed order could be well guided by other mechanisms, or by a cooperative working of several mechanisms, say by physical and biochemical ones. The fact that we observe the visible *implementation* of some effect does not mean that this implementation is the primary cause; it very well could be an *intermediate* instrument in the hands of the real, primary cause, of which we just are not aware. Unfortunately, Schrödinger disregards such a possibility, without reasons. He declares the living matter a special case, ignoring the previous scientific experience, and in particular the one acquired in physics and chemistry. One of the main pillars of this experience is that matter is governed by *hierarchical* structure of different laws of nature cooperatively acting at different scale levels. In many instances, the same law can act at different scale levels. Say, the Second Newton's law acts in the same way at a microbe spatial level, and also at a level of a planet. On the other hand, Archimedes' law and Hooke's law are applicable only at a spatial level where liquids and solid bodies appear, but not at a molecular level. The living world is as much a physical world as any other material thing the humankind is aware of. And so it also has to be governed by a combination of different physical laws. The idea of a "genetic code" as of a mechanism acting in molecular realm only, while defining everything at other spatial levels too, is in striking conflict with the arrangement of the physical world we know.

Cutting ties with a macro world and descending into a molecular realm, Schrödinger still adheres to physical principles: "*We must be prepared to find a new type of physical law prevailing in it. Or are we to term it a non-physical, not to say a super-physical, law? No. I do not think that. For the new principle that is involved is a genuinely physical one: it is, in my opinion, nothing else than the principle of quantum theory over again*."

On the other hand, Schrödinger does not exclude that Life is governed by not "super-physical law", but some more conventional law: "*We seem to arrive at the ridiculous conclusion that the clue to the understanding of life is that it is based on a pure mechanism, a 'clock-work' in the sense of Planck's paper, The conclusion is not ridiculous and is, in my opinion, not entirely wrong, but it has to be taken 'with a very big grain of salt*." The note "*the clue to the understanding of life is that it is based on a pure mechanism*" is a remarkable one; it presents a sound materialistic view of the world confirmed by practical experience of human history. It is clear from his paper that the only "rationale" behind his doubts expressed in the above quote is that he is rather subjectively inclined to assign the workings of such a physical law to *molecular* structures in cells of living organisms, without considering other quite plausible alternatives.

In any case, even such a strong proponent of a biomolecular paradigm of life origin and development, and to some extent its founder, could not dismiss the possibility that Life is governed by some 'ordinary' or special *physical* laws.

### 1.2. *Cooperative working of physical and biochemical growth mechanisms*

Such physical mechanisms will be presented in this article, thus introducing a more coherent and realistic paradigm of life development as a phenomenon, universally governed by *cooperative* workings of physical and biochemical mechanisms. The physical mechanism imposes macro-constraints, to which biochemical mechanisms have to comply. This cooperative working is not on the surface - its implementation is often complex, with feedback loops and lots of evolutionary adaptations affected by numerous and ever changing factors, but it's there, at the core of everything what is happening with living matter. One can see how biochemical reactions in a living organism change, without ever thinking why composition of biochemical reaction changes, what "engine" propels such ordered alternations. The current answer in biology is that this is done by some mysterious 'genetic switches', which nobody really knows, what does it mean - successive chains of biochemical reactions, epigenetic mechanisms and its changes, certain sequences in DNA, or all these things together, or maybe something else?

Physical mechanisms, presented in this paper, affect growth and reproduction at a cellular, organ, system and whole organism levels. We consider their application to modeling growth of unicellular organisms and understanding their division mechanisms, while referring to previously published articles considering modeling growth at the next spatial level - growth of organs. Knowledge of why and how cells grow and reproduce, what kind of fundamental mechanisms so universally and persistently govern cellular processes, is of great importance both for practical needs (medical, crop production, growth and productivity of domesticated animals, biotechnological, etc.), and scientific studies opening new areas for applications and explorations. Based on the obtained results, we eventually suggest to append the present view of biological phenomena as an overwhelmingly biochemical paradigm with certain physical mechanisms, thus moving in the direction of a *physical-biochemical* paradigm of Life. The remarkable thing, which adds much validity to our suggestions, is that this paradigm includes *all* presently known biochemical growth and division mechanisms as its inherent part, without a single conflict.

## 2. Present view of cellular growth and division mechanisms

Most studies on the subject of cell growth and division explore biochemical mechanisms, representing chains of biochemical reactions, implementing transitions through successive growth and division phases. Examples can be found in Refs. 3-8.

### 2.1. *“Timing” and “sizing” hypotheses*

Another direction of research was inspired by ideas to find *systemic* level mechanisms responsible for the cell growth and division. In his review[9], a prominent biologist Professor Mitchison says with regret: "*It would be satisfying if the main parameters of cell cycle growth had been established in the earlier work. Not surprisingly, however, there were still major uncertainties left when people moved from this field to the reductionist approaches of molecular biology*." Review Ref. 7 also accentuates specific properties of cellular growth and reproduction, which cannot be resolved exclusively at a biomolecular level.

Many concepts with regard to general growth and division mechanisms were proposed. Such are the "sizing" and "timing" hypotheses, claiming accordingly the priority of a cell size and of a certain time as primary factors defining the cell cycle progression. Professor P. Fantes[10] found experimentally that actually both "sizing" and "timing" homeostasis takes place, to a certain degree.

"Sizing" concept is represented by different, often conflicting, views. For instance, in Refs. 11, 12, the authors use an absolute size. Ref. 13 suggests that the cell cycle is driven by the "constant size extension". For the bacteria *Escherichia coli* and *Caulobacter crescentus* they infer that these bacteria "*achieve cell size homeostasis by growing on average the same amount between divisions, irrespective of cell length at birth*". The "constant size extension", in fact, is not constant, but noticeably varies. The authors acknowledge: "*The constant extension mechanism does not need to be precise, with experimental CVΔL of 19–26%.*", where 'CV' means standard deviation/mean.

Ref. 14 proposed complex relationship between the size and cell cycle in the form of a "noisy map". They say: "*noisy linear map implements a negative feedback on cell-size control: a cell with a larger initial size tends to divide earlier, whereas one with a smaller initial size tends to divide later*." However, their inference does not agree with the "constant size extension" suggestion (for instance, such are the results shown in Extended Data Figures 2 and 10 in Ref. 13, neither it agrees with experimental data across wider variety of organisms.

In Ref. 15, the authors came to a conclusion that "*The size of the cell at division is proportional to the initial size of the cell*", studying *Caulobacter crescentus* cells. This result contradicts both the "noise map" and the "constant size extension" hypotheses, and has no experimental confirmation for other organisms.

Authors of Ref. 16 disagree with all the above propositions, but think that, at least for *E. coli*, "*size control is effected by changes in the doubling time, rather than in the single-cell elongation rate*" and "*the current size is not the only variable controlling cell division, but the time spent in the cell cycle appears to play a role*".

In other words, neither the "sizing" hypotheses, nor the "timing" ones, nor combinations of both, could provide convincing proofs of universality of found relationships and explain all known observational facts related to a cell cycle control. However, the cited and many other works assume that there *should be* such controlling mechanisms, of which the experimentally obtained stable cell size distributions (including the ones in the aforementioned works) could be considered as indirect evidence. What is important to us is the undeniable fact that many people intuitively feel that such controlling mechanisms based on sensing cellular macro parameters (macro compared to a molecular level) exist. They just could not find such mechanisms yet. On the other hand, the implicit underlying assumption in these works remains the same - at the core, these "laws" are defined by some biochemical mechanisms.

More such works continue to appear, some claiming discoveries of universal and general laws of growth, which on closer look present rather particular relationships for certain organisms.

### 2.2. *Studies of growth mechanisms acting at higher than molecular level*

Note that the physical growth mechanisms and their interaction with biochemical machinery, which are considered in this article, explain *all* known to the author experimental results, reported in the literature for different organisms and different growth and division conditions.

From the perspective of scientific methodology, the Nature's laws and fundamental mechanisms do not work in such a simple way as the "sizing" or "timing" hypotheses assume. Fundamental mechanisms of Nature, at the least, are:

(1) Universal;

(2) Optimal; in the sense that from all possibilities their description requires the least possible number, and of the most fundamental, values, which all have to interrelate through the described by them phenomenon and to be adherent to the whole class of considered phenomena; if this is a mathematical description, then all these fundamental parameters have indispensable and irreplaceable roles;

(3) Provide the most possible *stability* of described phenomena without jeopardizing the scope of applicability, which has to include *all* such phenomena;

(4) Include parameters, which are both necessary and sufficient for the description of *any* phenomenon belonging to the problem domain;

(5) In the limit, when applied to previously studied *particular* scenarios, they have to produce the same earlier validated knowledge and results.

The most known and illustrative examples of laws of Nature, exhibiting these characteristics, are the laws of classical mechanics, electricity, thermodynamics. Ref. 17 presents more considerations, and in particular shows how the world's stability would suffer if Ohm's law changes just slightly.

Prominent scholar D'Arcy W. Thompson, in his prolific book "On Growth and Form"[18] presents considerations, supported by experiments, why there should be mechanisms, acting at higher than molecular levels, responsible for the growth of living organisms. A book "Life's other secret"[19] presents similar ideas and supporting proofs that the true "Secret of Life" is not in biomolecular mechanisms, and in DNA in particular. It also narrates that Crick, the discoverer of the double helix structure of DNA, once said in a pub, that "*DNA is not the secret of Life!*". Given the scientific rigour and thoroughness of the book, the citation almost surely presents genuine Crick's words; and if one thinks for a moment, this is not a phrase that some fiction writer could introduce, it could come *only* from a real life. Given numerous, and often outstanding, Crick's scientific achievements in different areas, as well as his personality, this is an opinion we should take seriously indeed.

Unlike the models of growth mechanisms reviewed in subsection 2.1, which do not satisfy above requirements for general laws and mechanisms, the *physical* mechanisms we introduce satisfy to *all* above criteria. These mechanisms, based on supporting experimental data, appears to be influential players in the growth and reproduction of cells[19-21], tissues, organs[22,23] and whole multicellular organisms. The presented mechanisms (a) *without a single exception* reconcile *all* known facts about cellular growth and division; (b) predicted certain growth and reproduction effects, which found experimental confirmations; (c) are seamlessly integrated with a cellular biochemical machinery; in fact, both physical and biochemical mechanisms work in tight cooperation, in which the presented physical mechanisms impose macro constraints the biochemical machinery complies with, and this is how the cell cycle progresses.

### 2.3. *Forces shaping biological phenomena*

The earlier discussed biochemical paradigm of Life assumes that life cycle of living organisms is coded in genes in the sense of successive binary switches (recall a quote from Ref. 2 n Introduction). The known workings of biochemical mechanisms, on the other hand, by no means exclude mechanisms of other nature, which could act at other spatial scale levels. In fact, at a closer unbiased look the biochemical mechanisms rather *appeal* for some "external" management, from higher scale levels. Such an arrangement is inherent to a physical world, when a multitude of different mechanisms, acting at different scale levels, shapes the same phenomenon. In this regard, living organisms represent an uninterrupted continuation of an inorganic world (such, Tobacco virus self-assembles in the presence of certain *inorganic* substances[19]). The book by Lane[25] presents hypotheses how *inorganic* matter could eventually produce living organisms in hydrothermal vents. By and large, there are no fundamental reasons that such a multifaceted phenomenon as Life, including the origin, life cycle and evolution of living species, should be defined entirely and exclusively by biochemical mechanisms alone. It is intuitively clear that the *objective* causes which led to appearance of living organisms existed *before* the biochemical mechanisms, and DNA in particular, were created. These "founding" mechanisms belong to an inorganic world, *from which* Life originated. It is much due to *their* action that living organisms originated and progressed through their evolutionary paths. Then, why the action of all these forces belonging to inorganic world had to stop and disconnect (as Schrödinger suggested) after the Life origin? For instance, DNA of an evolutionarily developed single cell could not contain everything needed for a multicellular structure, like a balanced growth of organs and systems in multicellular organisms. There should be other forces of nature, which took care of such tasks at appropriate scale levels – of course, besides the workings of biochemical machinery.

The well defined set of cell shapes, how did it happen? Was it only a random play of chemical reactions? Very unlikely, given that the optimal functionality of microbes and other microorganisms is supported, besides other macro-characteristics, also by certain geometrical shapes, like rods, spheres. What about the level of tissues, organs, systems, whole multicellular organisms? Should we still assume that these multi-scale constructions are managed from a molecular level? Maybe the shape of living organisms is also defined by some unknown mechanisms at higher than molecular level, in the same way as motion of planets, composed of innumerable number of molecules, is defined by Newtonian mechanics? Why not? This is how the physical world we know is arranged, and the living creatures present an inherent part of the *physical* world. It is just the *belief* in ultimate power of biochemical mechanisms alone, which separates us from such a step to a more multidimensional and comprehensive understanding of life phenomenon.

## 3. Physical growth mechanisms

### 3.1. *Introduction of a physical growth mechanism. The growth equation*

Here, we introduce and use the discovered growth mechanism for modeling growth and explaining division mechanisms of unicellular organisms. First, it was introduced in Refs. 26, 27, with the following advancements and applications in Refs. 20-24, 28-34 and other publications. The principal role of the main discovered mechanism is that it *uniquely* distributes nutrients, acquired by an organism, between the biomass synthesis and maintenance needs. In other words, using the mathematical representation of this general growth mechanism, the growth equation, we can find how much nutrients are used for biomass synthesis, and how much nutrients go to organism's maintenance needs, *at each moment of organism's life cycle*. The implications such knowledge provides are of fundamental value for biology and related disciplines. For instance, one of such important consequences is that the composition of biochemical reactions in both the entire organism and its constituents is *directly tied to the amount of produced biomass*.

Understanding this mechanism is rather difficult, for several reasons, such as its generality and non-obvious omnipresence in nature - for instance, in plants, because of their complex nutrients supply and waste removal structure. However, the greatest challenge, in the author's view, stems from the need to accept a *new view* that Life, besides biochemical mechanisms, is also governed by *physical* laws, acting at different scale levels, thus merging physical, biochemical, biological and other possible mechanisms into a single coherent concept of organic life.

One of the main physical phenomenon, underlying the general growth mechanism, is a conflict between the slower increasing abilities of the *surface* to supply nutrients and the nutritional needs of faster growing *volume*. However, in nature, this conflict is resolved not in *absolute*, as many researches assumed before, but *in relative* and *dimensionless* transformed form. The actual arrangement is more complicated and more elegant; it provides much greater flexibility, adaptability, stability and optimality than the surface-to-volume conflict in *absolute* values would allow. This is why all previous explanations involving size of organisms, or growth time, or both did not succeed.

Simplifying the matter for explanation, we can think of a spherical cell growing in three dimensions. Its surface increases proportionally to *square* of a radius (by four times for the radius's increase by two times), while volume increases proportionally to the *cube* of the radius (in our data, by eight times). If the nutrient supply per unit surface remains the same, that would mean that the unit volume of a grown cell will obtain twice as less nutrients (8/4=2). Organisms normally compensate for that, by increasing nutrient influx through the surface during growth. However, since the cubic function increases faster than the quadratic one, and the nutrient supply through the surface is *principally* restricted (by the environment and/or by the cell's membrane capacity), at some point, the surface inevitably won't be able supplying the same amount of nutrient per unit volume, and so the unit volume won't be able to function as before; then, the organism needs to do something, either "inventing" new mechanisms and moving to the next developmental phase, or to stop growth. In a nutshell, this is what limits the size of a cell for the given evolutionary formed metabolic mechanisms and available nutrient supply. What we described is just a rough description of the idea. Its realization in Nature is much more elaborate and elegant. The real mechanism senses not the *absolute*, but the *relative* values, and the *principle parameter* is defined not as a *ratio*, but a *ratio of ratios,* as we will find soon. The second distinction is that this is not a *biochemical*, *molecular* mechanism, working at molecular level, but a *physical* mechanism, acting at cellular and above spatial levels.

Of course, other factors of lesser influence could modulate the process too. Transportation expenditures also take a toll; the longer the communication routes, the more nutrients are required for transportation, and the less remains for other activities[22,35]. Similarly, the same conflict between relative volume and relative surface takes place for one- and two-dimensional growth. Even if nutrients are supplied through a stem (like in an apple), they are still distributed through the surface (internal surface, in this case), so that the relative surface - relative volume conflict is still there. The central location of a seed-bag in fruits, especially in the ones with short vegetation periods, besides other functions also creates an initial surface from which nutrients start distributing towards periphery.

One of the important consequences of such a resolution of the *relative* surface – *relative* volume conflict is that the fraction of nutrients used for biomass production is a value, which is uniquely defined by input growth parameters, first of all by geometrical characteristics. The rest of nutrients is used for maintenance needs.

The general growth mechanism explains why cells of the same species can grow large and small, depending on different factors. That's because the growth and division mechanisms, which act in large and small cells, are *the same*, and trigger successive growth phases and division at *the same* values of certain parameters regardless of the size and growth time (cells trapped in stones during volcano eruptions may have cell cycle measured in years, because of few nutrients).

The growth equation for a simple growth scenario, when nutrients are acquired through the cell surface, is as follows.

$$p_c(X)\mathrm{dV(X,t)}=\mathrm{k}(t) \times S \times \left(\frac{R_S}{R_V} - 1\right)\mathrm{dt} \qquad (1)$$

where $X$ represents a spatial coordinate, $p_c$ is the density of the cell (units of measure $\mathrm{kg} \cdot m^{-3}$), $t$ is time in *sec*, $k$ is a specific nutrient influx (amount of nutrient per unit surface per unit time) measured in $\mathrm{kg} \cdot m^{-2} \cdot \mathrm{sec}^{-1}$, $S$ is the total surface (in $m^2$) nutrients are acquired through; $V$ is volume (in $m^3$).

The left part of Eq. (1) is the mass increment. The right part is the product of the total nutrient influx ($K$) through the surface (the term $k(t) \times S$, so that $K = kS$), by a dimensionless parameter $(R_S/R_V - 1)$, called the *growth ratio* ($G$) (it includes that same *ratio of ratios* mentioned above); the growth ratio defines which *fraction* of the total nutrient influx is used for biomass production. Thus, the right part represents the *amount* of nutrients used for biomass synthesis in a time period *dt*.

The most important parameter in Eq. (1) is the growth ratio *G*. Why the nutrient distribution between the biomass synthesis and maintenance in nature has to be so definitive? The answer is this. The primary evolutionary goal of any living organism is its successful and *fast* reproduction (which assumes fast increase of biomass). No reproduction - no organism. For that purpose, the organism must use acquired resources *optimally*. If the biomass synthesis is non-optimal - for instance, too slow, then the reproduction process is in jeopardy. If non-optimal insufficient amount of nutrients is directed to maintenance, then the organism won't be able to produce biomass fast enough, and then the reproduction will be delayed too. So, there *should be* some optimal distribution of nutrients between maintenance needs and biomass synthesis, which provides the *fastest* rate of biomass production. Thus, nature, as is the case with its other fundamental mechanisms and laws (which is especially evident and illustrative in case of mechanical laws - recall Hamiltonian and generalized coordinates), goes *on an optimal path*, securing the fastest reproduction time for the given conditions, doing this through optimal resolution of the relative surface – relative volume conflict.

What is also extremely important, this optimal path provides *the greatest stability possible* for a given phenomenon[17,20]. (The fundamental stability of the world we know is the *consequence* of such optimality of laws of nature, by the way.) This optimality is tied to a certain *geometrical* form. *All* organisms do have some geometrical form, which is the universal base, and this is why such a nutrient distribution is *universal* for all species and their constituents, from the cellular level to organs to whole multicellular organisms. The growth ratio is a mathematical representation of this optimal distribution of nutrients, implemented in Nature.

In physics, the *same* principle of maximum stability due to optimality is behind the facts that acceleration of a body is directly proportional to applied force and inversely proportional to mass, or electric current in a circuit is directly proportional to applied voltage and inversely proportional to resistance. It was shown in Ref. 17 that when relationships between fundamental parameters deviate from such an optimum, the world which we know would be unlikely to exist, because of the inherent *instability* brought by these apparently minor changes. The growth ratio and the general growth mechanism are from the *same* category of fundamental parameters and relationships between them.

Note that the growth ratio and the growth equation were discovered *heuristically*, which was *the only way* for discovering fundamental growth parameters and relationships between them[20], since there was nothing to derive them from. (Recall famous "Eureka!" by Archimedes, add Descartes' clarity, and this will give an idea how the growth equation *emerged*.) It is defined as follows. Suppose that a cell can grow to a maximum volume $V_{\max}$, which has a maximum surface $S_{\max}=S(V_{\max})$. Then, the dimensionless parameters - a relative surface $R_S$ and a relative volume $R_V$, are as follows:

$$R_S = S/S_{\max} \quad (2)$$

$$R_V = V/V_{\max} \quad (3)$$

The dimensionless growth ratio *G* is defined as follows:

$$G=\frac{R_S}{R_V}-1 \quad (4)$$

As a function of volume, the growth ratio monotonically decreases when the organism's volume increases. According to Eq. (1), it means that the more the organism grows, the less nutrients are available for biomass synthesis, and more nutrients are used for maintenance. This is understandable, since the growing biomass requires more and more nutrients for maintenance. Eventually, this conflict stops the growth.

The maximum size can change during growth depending on nutrients availability and other parameters. Such, a cell that begins to grow in an environment with low nutrient content, is destined to have a smaller final size. However, if, at some phase of growth, the environment is enriched in nutrients, then the cell grows bigger, which experimental observations confirm[13,36,37]. In many instances, the maximum size can be known upfront, if all growth conditions are defined at the beginning and do not change unpredictably later. Otherwise, the maximum size can change. This mathematical specific of the growth equation does not mean that it has some defects or it is of approximate nature. This is just an adequate mathematical description of the growth phenomena in nature, when the change of parameters during the growth *principally* can alter the final size of a grown organism – this is a fact proven experimentally.

In modelling, the possible variability of the maximum possible volume can be addressed by adding the dependence of its value from other parameters; for instance, from nutrient influx, temperature. (This, of course, mathematically will significantly complicate the growth equation and its solution, but won't make finding the solution *principally* impossible.) Below we will discover that in certain types of growth scenarios knowing the maximum volume is not required.

### 3.2. *Finding nutrient influx*

The next important parameter in Eq. (1) is nutrient influx $k$, the amount of nutrients per unit surface per unit time. We already mentioned that the product $k \times S$ represents the *total* nutrient influx $K$, so that $K = k \times S$ . This fact reflects the property of the growth equation that it does not matter which way the nutrient influx was acquired. Such, in a growing budding yeast part of nutrients comes from the mother cell; in an apple nutrients come through the fruit's stem. Thus, to solve Eq. (1), we can either find $k$ and $S$ separately, or find their product $k \times S$ as the *total nutrient influx K*. Unfortunately, we have no reliable data for the nutrient influx $k$ per unit surface, how it changes with the growth. However, we do have results of experimental studies related to nutrient consumption by cellular *volume* with regard to different cellular components, so that using this knowledge we can find the total nutrient consumption $K = k \times S$ and substitute it into Eq. (1) instead of the term $k \times S$.

In Ref. 22, 35, the amount of nutrients required for cellular transportation depending on the shape of cells was found. In Ref. 22 the overall amount of nutrients required for the growth of *S. pombe* and *Amoeba* was obtained. It was discovered in Refs. 37, 38 that in some elongated cells, like *E. coli*, *S. cerevisiae*, the rate of RNA synthesis is twice the rate of protein synthesis. Taking into account this double rate of nutrient consumption for RNA synthesis, we can write for the nutrient influx $K_{\min}$ required for biomass synthesis and maintenance (without transportation costs) the following.

$$K_{\min}(v)=\mathrm{N}\left(C_p^s v+\mathrm{C}_r^s v^2\right) \tag{5}$$

Here, $C_r^s$ and $C_p^s$ are fractions of nutrient influx required for RNA and protein synthesis; $v$ is the relative increase of organism's volume (the ratio of the current volume to the volume at the beginning of growth, that is $v=V/V_b$, and consequently $v\geq 1$) ; $N$ is a constant.

For an elongating cylinder-like cell, whose diameter remains constant, volume is proportional to the relative increase of length $L$ (ratio of the current length to the beginning length). Using the same consideration as in Refs. 22, 34 about proportionality of transportation costs to the traveled distance, and substituting $K_{\min}$ from Eq. (5), we obtain an equation for the total nutrient influx.

$$\mathrm{dK}(L)=\mathrm{C}_t K_{\min}(L)\mathrm{dL} \tag{6}$$

where $C_t$ is a constant.
Solving (6), we find

$$K(L)=\mathrm{N}_1 C_t\left((1/2)C_p^s L^2+(1/3)C_r^s L^3\right)=\mathrm{A}\left(C_p L^2+\mathrm{C}_r L^3\right) \tag{7}$$

where $N_1$ is a constant; $\mathrm{A}=\mathrm{N}_1\, C_t/2$; $C_p=\mathrm{C}_p^s$; $C_r=(2/3)C_r^s$.

Later, we will actually model elongated cells by a cylinder with hemispheres at the ends, while (6) does not account for hemispheres. However, volume of a cylindrical part in such real organisms is greater than the one of both hemispheres, and this is why this approximation provides sufficient accuracy for our modeling purposes. Besides, when an elongated microorganism grows, volume of a cylindrical part increases, while volume of hemispheres remains constant, which further increases the accuracy.

Similarly, we can find the required total nutrient influx $K$ for a disk and a sphere. Below, we assume that the rate of RNA synthesis is twice the rate of protein synthesis. A disk grows in two dimensions (height remains constant); a sphere increases proportionally in three dimensions.

$$K(v)_{\text{disk}}=A_d\left(C_p v^{3/2}+\mathrm{C}_r v^{5/2}\right) \tag{8}$$

$$K(v)_{\text{sph}}=A_s\left(C_p v^{4/3}+\mathrm{C}_r v^{7/3}\right) \tag{9}$$

where $A_d$ and $A_s$ are constants.

Obtaining analytical solutions as (7) - (9) is not always possible. Such, there is no analytical solution for an elongating ellipsoid (used for modeling growth of *S. cerevisiae*). In this case, the following growth equation should be solved numerically[20,32].

$$p_c \mathrm{dV}(r)=\mathrm{k}_{\min}(V(r))\times(r/r_b)\times S(r)\times\left(\frac{R_S}{R_V}-1\right)\mathrm{dt} \tag{10}$$

Here, $r_b$ is the beginning radius in the same direction of growth, which is defined by a radius-vector $\boldsymbol{r}$.

## 4. Growth model of *Amoeba* and its division mechanism

### 4.1. *Amoeba's growth model on the basis of the general growth mechanism*

We will start from a simpler growth scenario, which is implemented in *Amoeba*. *Amoeba* might be not the oldest organism, but there are no reasons why it cannot use a primordial growth and division mechanism, if it serves the purpose; in the same way, we still use an ancient tool, a hammer, because it is adequate for our tasks. (As a side note, an interesting consideration is presented in Ref. 25 that the ancestor's root of eukaryotes, including protists, like *Amoeba*, could be *much older* than it is presently assumed. If this is so (and the general growth mechanism, in fact, indirectly supports this finding), then the basic growth and reproduction mechanisms could be the only option for *Amoeba's* ancestor, and more sophisticated mechanisms for other organisms were built later *on top* of these basic mechanisms.) Experimental data for *Amoeba* are from an exemplary study presented in Ref. 39.

*Note*: It is very important for the reader to understand how the rigorousness of experimental validation of the growth equation and of the growth mechanism was achieved. When people look at presented curves and experimental points, they automatically assume that the curves are just the fitting curves associated with some statistical model. Nobody is to blame for such an (invalid in this case) assumption, because this is what almost all such articles present. In this regard, the approach used in this paper is *very different* and much more *robust,* and it is *not* based on a data fitting procedure. In our case, the *entire* theoretical growth curve is calculated deterministically by the growth equation *alone.* From experimental data, we use the point at which growth ends and division begins. This is a very reasonable input to have, since, as we know, cells variate in size. All other inputs, with rare exception, follow from theoretical considerations, associated with the growth equation and the growth mechanism; no ad hoc assumptions, usually routinely used in growth models, were added. One should agree that such an approach is a very robust validation of any concept and / or of mathematical equation describing natural phenomenon. The same validation procedure is used for other considered organisms, that is we always *first* compute the theoretical growth curve using the growth equation, and *only after that* compare it to the whole set of experimental points.

Let us define parameters of the growth equation.

*Density and mass calculation*. For all considered microorganisms, we assume the density to be constant during the growth and equal to 1 $g \cdot \text{cm}^{-3}$. This assumption is a reasonable approximation[39]. The mass of a grown *Amoeba* used in calculations corresponds to the last experimental measurement.

*Maximum possible volume*. A parameter "spare growth capacity" (SGC) was introduced for the characterization of the maximum possible volume in Refs. 20,31. It is defined as $\text{SGC}=1 - V_d/V_f$. Here, $V_d$ is the volume when cell divides; $V_f$ is the maximum possible volume, which the growth curve asymptotically approaches. For available experimental data, for *Amoeba* SGC value was in the range 1.0 - 2.8%. Two *Amoebas* did not divide and, *indeed* (!), increased their mass by about 2% after missing division, so that SGC is a *real* value - this is important to note. In calculations, we used the maximum possible volume, which exceeded the last measurement by 2%.

*Geometrical form*. *Amoeba* is modeled by a disk whose height $H$ is equal to the initial disk radius $R_b$. The maximum possible disk radius is $R_0$, the current radius of a growing disk is $R$. Such a model was chosen based on analysis of *Amoeba's* images from different sources, which indicate rather two-dimensional increase of this species. (A more sophisticated pinion-like form, accounting for *Amoeba's* pseudopods and which was also studied, produced close results.) Substituting the above parameters into (2) - (4), we obtain:

$$R_{\mathrm{Sd}} = \frac{R(\mathrm{R+H})}{R_0(R_0+\mathrm{H})};\ R_{\mathrm{Vd}} = \frac{R^2}{R_0^2};\ G_d = \frac{R_{\mathrm{Sd}}}{R_{\mathrm{Vd}}} - 1 = \frac{R_0(\mathrm{R+H})}{R(R_0+\mathrm{H})} - 1 \qquad (11)$$

where index *'d'* denotes 'disk'. Here, we use a capital letter '*R*' both for surface and volume ratios $R_S$ and $R_V$ , and also for radii. So please do not be confused and pay attention to indexes. Once the ratios are defined though, as it was done in (11), we work with radii only.

*Nutrient influx.* The rate of nutrient consumption for RNA and protein synthesis are assumed to be the same for *Amoeba*, which follows from Refs. 20,22,31. This transforms Eq. (8) into

$$K(v)_{disk} = A_d v^{3/2}(C_p + C_r) = A_d[(R/R_b)^2]^{3/2}(C_p + C_r) = A_d(R/R_b)^3(C_p + C_r) \quad (12)$$

*Model verification*. Using the above parameters, as it was said already, we ***first*** computed the growth curves, and only ***then*** compared them with experimental dependencies. So, this comparison ***is not*** a data fitting procedure, but actually a *principally* much more rigorous verification of the model's adequacy.

*Solution of the growth equation*. Substituting the above parameters into Eq. (1), and remembering that the term $k \times S$ is substituted by the term $K(v)_{\mathrm{disk}}$ from (12), we obtain the following differential equation.

$$2\pi p RHdR = A_d(R/R_b)^3(C_p + C_r)[R_0(R+H)/(R(R_0+H)) - 1]d\tau$$

which can be transformed to

$$2\pi p dR = A_d(C_p + C_r)R(R_0 - R)/[(R_0 + H)R_b^3]d\tau$$

and then to a more explicit form with separated variables as follows

$$d\tau = \frac{2\pi p(R_0 + H)R_b^3}{A_d(C_p + C_r)}\left(\frac{dR}{R(R_0 - R)}\right) \qquad (13)$$

The expression with variable $R$ in the right brackets can be presented as

$$\frac{dR}{R(R_0 - R)} = \frac{1}{R_0}\left(\frac{1}{R} + \frac{1}{R_0 - R}\right)dR$$

Then we should solve the integral equation

$$\int_0^t d\tau = \frac{2\pi p(R_0 + H)R_b^3}{A_d(C_p + C_r)R_0}\int_{R_b}^{r}\left(\frac{1}{R} + \frac{1}{R_0 - R}\right)dR$$

where $r$ is the current radius of a disk at time $t$ (at start of growth $t = 0$).

Solving this equation, we obtain

$$t = \frac{2\pi p(R_0 + H)R_b^3}{A_d(C_p + C_r)R_0}\ln\left(\frac{r(R_0 - R_b)}{R_b(R_0 - r)}\right) \qquad (14)$$

Here, we used the fact that

$$\int\left(\frac{1}{R}+\frac{1}{R_0-R}\right)dR=\ln R-\ln(R_0-R)+const$$

Certainly, we can use (14) to draw the growth curve, and that was the approach used in the earlier articles. However, it would be better to find a direct analytical solution for the radius *r* as a function of time *t*. The solution, indeed, is an absolutely remarkable one from the perspective of criteria of validation of scientific truths, derived from the principles of philosophy and natural philosophy. Let us denote

$$c_0=\frac{2\pi p(R_0+H)R_b^3}{A_d(C_p+C_r)R_0}$$

Then the solution of (14) with regard to *r* as a function of time *t* is as follows.

$$r=\frac{R_0\exp(t/c_0)}{(R_0/R_b-1)+\exp(t/c_0)} \qquad (15)$$

The remarkable thing about Eq. (15) is that this is a *generalization* of a solution

$$P(t)=\frac{\exp(t)}{1+\exp(t)}$$

of the well known logistic equation

$$\frac{\mathrm{dP}}{\mathrm{dt}}=(1-P(t))P(t),$$

where *P* is the population quantity. However, our solution Eq. (15) has been obtained *independently*, on very different grounds. Unlike the classic solution, which requires adding constant coefficients using ad hoc considerations, Eq. (15) produces *all* coefficients *naturally*, as functions of the model's input parameters, which is a significant and *qualitative* advancement over the logistic equation. So, we obtained that the heuristically introduced growth equation produced a *generalized* solution, whose particular implementation produces solution of the known logistic equation. However, the logistic equation is *also* used for quantitative modeling growth phenomena; not only for populations, for which purpose it was derived, but for many other growth scenarios. This fact should be considered as a remarkable verification result, a strong argument in favor of validity of the growth equation, according to criteria for validation of scientific truths; in particular that a more general theory (or equation) must converge to more particular cases earlier obtained for the same phenomenon (if available, of course, which is fortunately our case). Indeed, the well known and widely used logistic equation turned out to be a *particular* case of the *general growth equation* for a disk like growing organism. (Later we will see that logistic equation is also a particular case of the general growth equation for a sphere-like growing organism.) Finding such a connection creates a very strong support for the validity of the growth equation.

In fact, the obtained results cross the border into a realm of *natural philosophy*, in which the found relationships of *a general* and of *a particular* present high value and serve as a solid proof of validity of scientific hypotheses.

Note that the growth equation represents a *new equation of mathematical physics*, so that obtaining such an interesting and significant results is a good start for its mathematical explorations too.

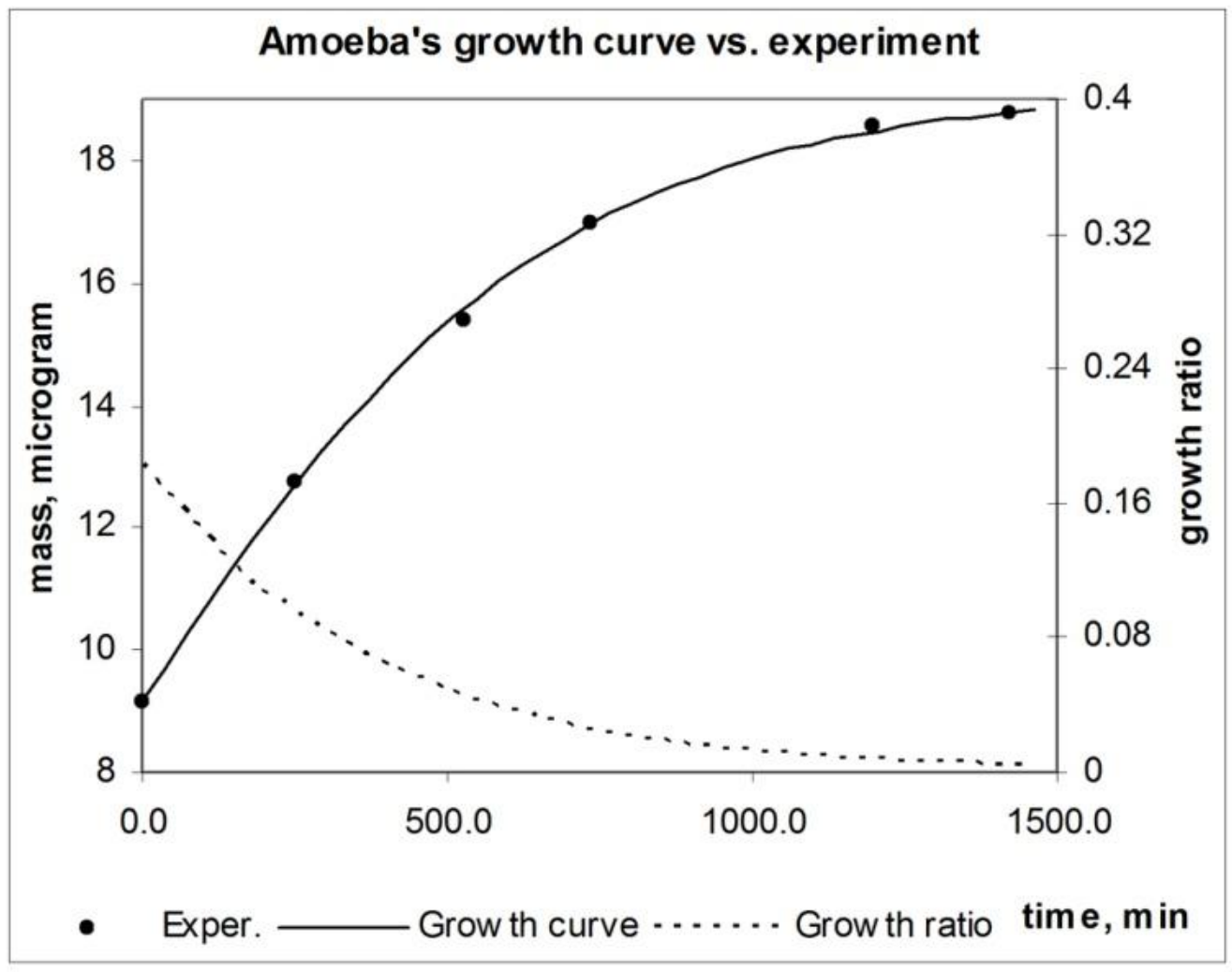

Fig. 1. Computed *Amoeba*'s growth curve versus experiment, and the computed growth ratio, depending on time. Experimental data are from Ref. 39.

Fig. 1 shows the computed growth curve for *Amoeba* versus experimental data from Ref. 39, and the corresponding growth ratio. We can see that the computed growth curve corresponds to experimental measurements very well. Comparison with other experiments shows slightly more dispersion of experimental points relative to the computed growth curves[20]. However, the experiment in Fig. 1 was chosen not for the least deviation from the computed growth curve, but for the *stability* of growth conditions compared to other experiments, in which nutrient influx was not so stable. If we could know the actual nutrient influx, we would compute the growth curves for other experiments more accurately too.

### 4.2. *Amoeba's division mechanism*

A continuous redistribution of nutrient influx between maintenance needs and biomass production, defined and enforced by the general growth mechanism, explains deceleration of the growth rate and subsequent stopping of growth. Indeed, growing biomass requires more nutrients for maintenance to support it, and so fewer nutrients are available for biomass synthesis. The decrease of the growth ratio during growth is a quantitative expression of this fact, in a mathematical form. This arrangement of the growth phenomena in Nature has far reaching implications. Here is why. Organismal biochemical machinery represents a *single unity*. There are no separate biochemical machinery for maintenance and for biomass production, but *all* biochemical reactions *interrelate*; they are arranged in such a way that output substances of previous reactions become inputs for the next. Success of methods of metabolic flux analysis is based *entirely* on this arrangement, when through such interdependencies, described by a system of stoichiometric equations, it is possible to *unambiguously* find how much of each substance participates in the biochemical interchange[31,40].

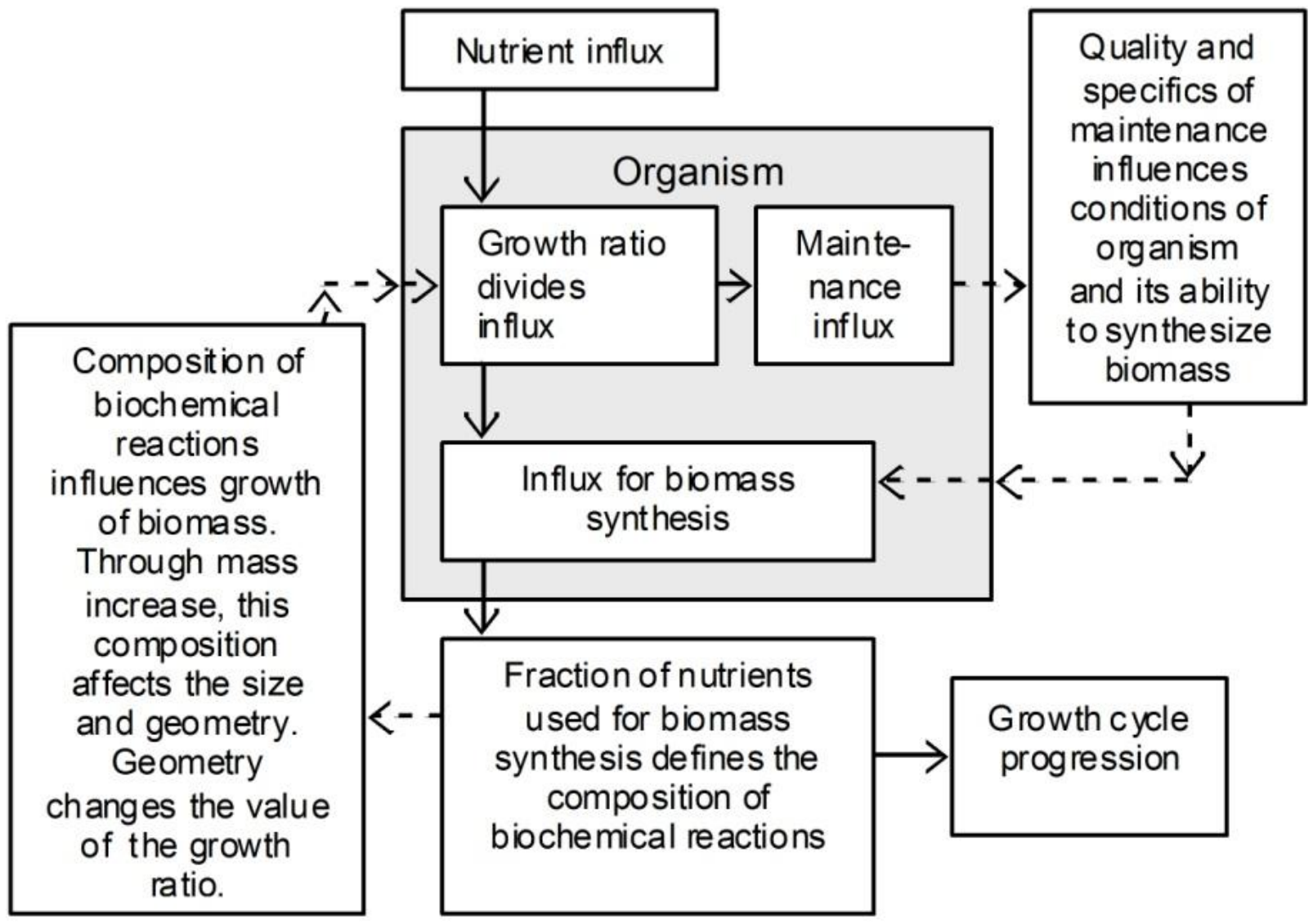

Fig. 2. Growth cycle regulation and progression defined by the general growth mechanism.

According to works on metabolic flux analysis, the solution of a system of stoichiometric equations produces *the most adequate* results when this solution is optimized for a *maximum amount of produced biomass*. Indeed, evolutionary development shaped the composition of biochemical reactions in the direction prioritizing fast reproduction; in other words, making the amount of produced biomass a *leading* parameter, to which the composition of biochemical reactions is tied. So, if the amount of synthesized biomass changes, the composition of biochemical reactions changes too. In which direction though? The answer is: In the direction securing successive transitions through the entire growth period, optimized for the *fastest* reproduction in the given conditions.

Before proceeding further, we should make the following side note. Nutrients are transformed to biomass by chemical reactions, for which the law of conservation of matter is fulfilled, so that the *mass of nutrients*, which are used for biomass synthesis, is equal to the mass of *synthesized biomass*. Therefore, in the following, it is legitimate using interchangeably these two values.

Let us reiterate how the growth and reproduction is regulated from the standpoint of the general growth mechanism using Fig. 2. In nature, a conflict between supplying abilities of the surface and the faster increasing demands of volume is resolved through optimization of nutrient distribution between biomass synthesis and maintenance. Quantitatively, this optimum is expressed as the value of the growth ratio, which is defined by geometry (that is logical, since the surface and volume are primary geometric characteristics inherent to all living organisms, and this is where nutrients reside and move). This way, through the growth ratio, the general growth mechanism imposes constraints on the fraction of nutrients that go to biomass production at each moment of growth. Biochemical mechanisms comply with this constraint, adjusting the amount of produced biomass to the amount of nutrient defined for that purpose by the growth ratio, by changing the composition of biochemical reactions. Such an arrangement *had to be* evolutionary developed by natural selection process, since organisms whose distribution of nutrients was closer to values defined by the growth ratio, grew faster and so did have evolutionary advantage. And the change of compositions of biochemical reactions in such organisms proceeded accordingly to synthesize at each moment the amount of biomass, close to the one defined by the growth ratio. Eventually, given enough evolutionary time, the change of compositions of biochemical reactions will adjust to values defined by the growth ratio very accurately. (This accuracy will be evident when we compare more experimental and theoretical growth dependencies.)

The increasing biomass requires more nutrients for maintenance, and so lesser fraction of nutrients is available for biomass synthesis. This continuous nutrient redistribution is guided (and accordingly reflected) in the value of the growth ratio, which monotonically decreases during the life cycle. Composition of biochemical reactions is tied to the *relative* amount of produced biomass (relative to the total amount of acquired nutrients, or, which is about the same, but a more sensitive characteristic, to the amount of nutrients used for maintenance). Thus, the changes in the relative amount of produced biomass, forced by changing geometry, change in the composition of biochemical reactions, so that the newer composition corresponds to the new relative amount of produced biomass. (In particular, such change of composition of biochemical reactions forced by the changed amount of produced biomass could be one of the main factors triggering cell specialization. However, this hypothesis requires further studies.)

*Irreversibility of a reproduction cycle*

We know that growth processes and organisms' life cycles are generally *irreversible*, including such phenomenon as aging of multicellular organisms. The described arrangement *explains* irreversibility during growth and reproduction. The decreasing relative amount of nutrients diverted to biomass production acts as a ratchet, preventing the current composition of biochemical reactions to revert to the previous state, when a greater relative amount of biomass was produced. For such a reversion to happen, the fraction of nutrients used for biomass production has to increase. However, the grown biomass already took for maintenance the part of nutrient influx, which was earlier used for biomass synthesis; and the entire biochemical machinery was adjusted accordingly. In order to increase the fraction of nutrients for the biomass production, the growth ratio has to be increased. It can happen through the size reduction or by substantial change of a geometrical form, which is not impossible, but would cause certain energetic, functional and developmental complications. If the said is true, then at least some simpler organisms might be able to "rejuvenate" through the decrease of biomass. Indeed, in Refs. 41,42, the authors acknowledged that by periodically resecting part of *Amoeba's* cytoplasm *it is possible* to indefinitely prevent it from entering division; in other words, making it practically immortal. By reducing *Amoeba's* size, the experimenter, in fact, increased the growth ratio. Led this to automatic adjustment of composition of biochemical reactions corresponding to a greater value of the growth ratio (in other words, to a new fraction of nutrients that could be used for biomass synthesis), or whether this led to freezing for some time the composition of biochemical reactions existing at the moment of resection, we don't know, but that could be found experimentally.

However, in general, the growth cycle is difficult to reverse for the reason explained above, and this is the price for the smooth and persistent proceeding through the entire growth and reproduction cycle. (The objection to the said above can be that there are cells that divide without growth. However, these cells reside within multicellular organisms, whose other parts increase their biomass and can send appropriate signals to other cells, forcing them to divide. The other possibility to start division without increase in size is to decrease the growth ratio to the point of division by changing geometry.)

The described division process is rather a backbone mechanism, which, as usual, can be modulated by nature-virtuoso in many ways, but these modifications, still, are built on top of this core mechanism.

Quantitatively, the leading role of the growth ratio (or, which is the same, of nutrient distribution between biomass production and maintenance) in the rate of change of the amount of synthesized biomass can be confirmed as follows. Let us rewrite Eq. (1) to explicitly show the amount of produced biomass $m_b$.

$$\mathrm{dm}_b = \mathrm{k(X,t)} \times S(X) \times \left(\frac{R_S}{R_V} - 1\right) \mathrm{dt} \qquad (16)$$

As we will see later, the specific nutrient influx $k$ for *Amoeba* at the end of growth changes little, as well as its volume and consequently the surface area. From the three terms in (16), the growth ratio G=$(R_S/R_V - 1)$ changes by far the quickest, so that the changes in the amount of produced biomass are defined mostly by changes of the growth ratio. It is important to understand that it is not the absolute, but the *relative* changes in the amount of produced biomass, which alter the composition of biochemical reactions. 50% decrease of small amount of produced biomass affects composition of biochemical reactions *more* than 25% decrease of a bigger amount of synthesized biomass.

It is clear that for the division mechanism to be efficient, it has to satisfy the following requirements:

(1) to be tied to the most important organismal characteristics;

(2) rate of change of these characteristics has to be *substantial* when approaching the division phase;

(3) for the same species (or maybe even for a class of species, having similar, at the core, growth and reproduction mechanisms) the values of these characteristics (or a characteristic) have to be *invariant* to all possible growth and reproduction scenarios (at least to be invariant with high accuracy).

The growth ratio satisfies all these criteria, while other competing parameters do not. For instance, the other candidate for the role of such a division trigger is often assumed to be the size of an organism. Fig. 3 presents graphs of a relative change of the growth ratio and the relative change of volume during the whole growth period for equal time intervals $\tau$, that is the values $(V(t+\tau) - V(t))/V(t)$ and $(G(t+\tau) - G(t))/G(t)$. We can see that relative changes of volume before the division are *substantially smaller,* than the relative changes of the growth ratio. Moreover, volume's relative change *decreases*, while in case of the growth ratio the relative change *remains constant* and its value is much greater (at sixty times at the end). Apparently, a triggering mechanism, which reacts to a *greater* and (even better) *increasing* parameter (which is the case for *S. pombe*, as we will see later), will work more reliably than a trigger reacting on a small and decreasing value.

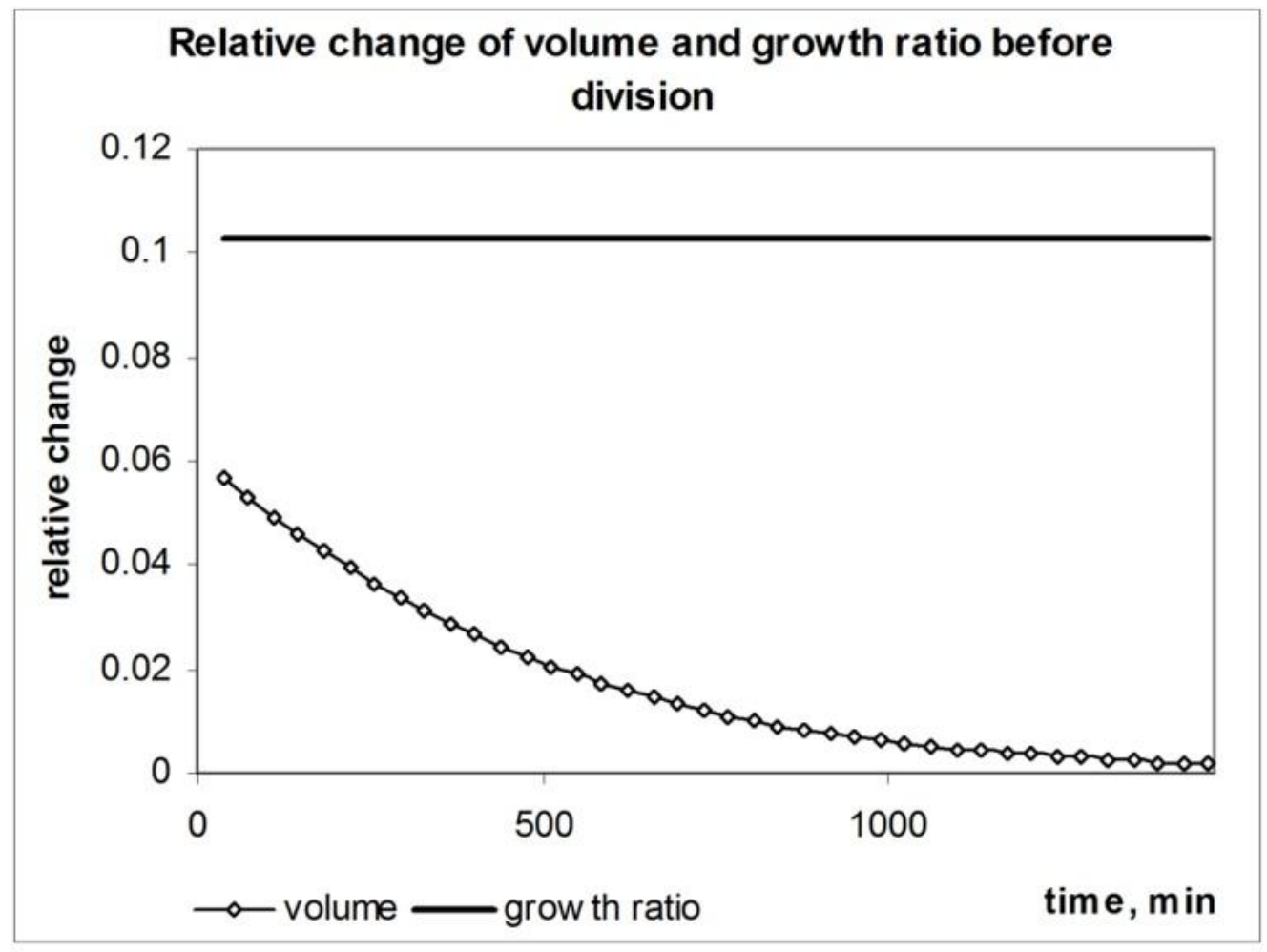

Fig. 3. Relative change of volume and growth ratio for *Amoeba* during the growth period.

The other important factor supporting our hypothesis about the role of the growth ratio as a major division trigger is this. Size of the same species varies a lot. Such, the length of a grown *S. pombe* can differ as much as four times; other cells also show wide range of grown sizes, at least tens of percent, depending on many factors, like nutrients availability, temperature, etc. In individual growth, cells also demonstrate wide variations of ratios between the ending and the initial sizes. Small *S. pombe* or *E. coli* can grow into a big cell, on par with the cells, which started growing being much bigger already. How some hypothetical size sensitive division mechanism could determine, at which size it has to start the division, given such *principal* variability of sizes relative to initial sizes, and also high variability of the initial sizes too, for the same species? Such a division mechanism just has no reference points to be tied too. On the other hand, the growth ratio is a well defined value for *any* growth scenario, and this value is directly tied to a composition of biochemical reactions – and to start a division it is namely such a change in composition of biochemical reactions is needed. For the same species with similar geometrical forms, it does not matter, a big cell or a small one, increases it by two or four times, the growth ratio will be changing similarly during the entire growth period, *always* reaching a certain value, which is *invariant to size*, corresponding to a division point. In other words, this division point corresponds to the same fraction (fraction of the total amount of nutrients) directed towards biomass production. In turn, it is *namely this* fraction, which defines and forces changes in the composition of biochemical reactions throughout the life cycle, until composition of biochemical reactions will change to the one corresponding to a division point. After which the division starts and proceeds till creation of a new cell. With the growth ratio as a division trigger, *all* known facts are explained, and *all* affecting factors are tied together. In case of size as a possible trigger of a division mechanism, on the contrary, we have conflicting considerations; and it just has no reference points a division mechanism can be tied too.

The invariance of value of the growth ratio, corresponding to division point, is indirectly confirmed by results from Ref. 14: "… *a cell with a larger initial size tends to divide earlier, whereas one with a smaller initial size tends to divide later*." Indeed, according to the general growth mechanism, large cells of the same species reach the threshold division value of the growth ratio earlier, since they start with a smaller growth ratio already and so it will decrease faster, while smaller cells generally start with a greater value of the growth ratio and need to grow - in relative terms - more in order to reach the same small value of the growth ratio corresponding to division. Of course, the growth ratio is defined by *geometry* at the first place, not by the size alone, but same type cells have similar forms, like cylinders or spheres, and so the close value of the growth ratio corresponding to the beginning of division.

So, the growth ratio in this competition for the cellular cycle control by far supersedes the organism's volume (and consequently other related to size absolute parameters). Thus, this is not the size of an organism, which triggers the division and enforces ordered changes of compositions of biochemical reactions through the growth cycle, but the *change* in the relative amount of *produced biomass*. And this change is tied to a *relative* size and to a geometrical form. (The geometrical form, as we will see later, is generally optimized for the greatest value of the growth ratio, which means the fastest production of biomass, except for the cases when the growth has to be suppressed, which is also done through such a change of geometry that decreases the growth ratio and consequently slows down the growth rate. Thus, geometry, the rate of growth and change of composition of biochemical reactions are all tied together through the growth ratio.)

This presented arrangement is certainly outside the mainstream biochemical biological paradigm, as well as outside the scope of numerous hypotheses about the growth mechanisms, but unlike other hypotheses, the general growth mechanism hypothesis supersedes all other hypotheses in many respects: It explains *all* known facts about growth and reproduction phenomenon, *without a single exception*, while other hypotheses have very limited application.

The general growth mechanism also allowed predicting new effects. Such were explanations why organisms have certain geometrical forms, like cylindrical or spherical ones; a theoretical discovery of a growth suppression mechanism based on change of a geometrical form[20,22,31]. The last effect later found experimental confirmation in cellularization of the syncytial blastoderm in Drosophila[43] and pigs' blastocysts[44]. These effects were discovered based on calculating growth time for different geometrical forms using the growth equation, and later we found experimental confirmation.

Unlike other hypotheses, the introduced *Amoeba's* growth and division models are supported by a mathematical apparatus, which produces results very accurately corresponding to experimental data. None of the other hypotheses about growth and reproduction mechanisms has such an adequate and universal mathematical apparatus, and passed such a robust verification based on strict scientific methodology. So, although the discovered answer to Life development problem resides in the area not many biologists expected it to be (indeed, we discovered that this is rather a classical *physical* mechanism working together with biochemical machinery), the discovery should not be discarded on that ground. Important scientific breakthroughs, like the military ones, originate in *unexpected* directions. If so many attempts made on the bitten grounds failed, maybe it worthwhile to look for another land of opportunities?

### 4.3. *Appearance of exponential function as a one more proof of validity of the growth equation*

A note about the constant value of the relative change of the growth ratio in Fig. 3. This is a surprising result. Its mathematical proof is as follows. Let us substitute the value of $r$ as a function of $t$ from (15) into the expression for the growth ratio $G$ in (11). (We should assume in (11) that $R = r$, due to the change of a variable when we did integration in Eq. (13).)

$$G(t) = \frac{(R_0/R_b - 1)H}{(R_0 + \mathrm{H})\exp(t/c_0)} = \frac{(R_0/R_b - 1)H}{(R_0 + \mathrm{H})}\exp(-\,t/c_0) \qquad (17)$$

Therefore, the growth ratio as a function of growth time is an *exponential function*. Recall that the growth ratio is the ratio of the *relative* surface to *relative* volume (Eqs. (2) and (3)), minus one, and *none* of these input values, of course, contains even a hint to exponential functions. Obtaining such an unexpected result with regard to relationships between the relative surface and relative volume in the given circumstances rather means that we found some important new geometrical property of the *real world* (which is, in the first place, a geometrical one).

The emergence of an exponential function is a remarkable result for the proof of validity of the growth equation. The first derivative of an exponential function is also an exponential function, so that the relative change of the growth ratio (its first derivative) is an exponential function too. For illustrative purposes, let us consider equal discrete time intervals $\tau$ though. Then, we can find the relative change of the growth ratio as follows.

$$\frac{G(\mathrm{t}+\tau) - G(t)}{G(t)} = \exp(-\,\tau/c_0) - 1 \qquad (18)$$

So, for equal time intervals the relative change of the growth ratio, indeed, remains *constant*. What is the meaning of this relationship in the real world? It is an interesting, very important and a *very natural* one. It means that at equal time intervals the amount of nutrients that is diverted to biomass production is reduced *by the same fraction* from the amount of biomass produced in the previous time interval. Suppose, at some point of growth we have a unit of synthesized biomass, and the fraction that goes to biomass production during the next time interval is 0.9 of the biomass produced in the previous interval. Then, we will have a sequence of produced biomass in each successive time interval as follows 1, 0.9, 0.81, 0.73, 0.656, ... ; each next biomass is obtained from the previous one by multiplying it by 0.9. Recall that exactly in the same way behave many other natural processes: they also contain exponential dependencies with the same natural number *e* as the basis, such as a decrease of atmospheric pressure with height, decrease of current in electrical circuits[17,20], attenuation of electromagnetic waves in absorbing media, etc. - mathematically, these are phenomena of the same class. There are many such natural processes defined by fundamental laws of Nature. So, we can say with certainty that the *heuristic* growth equation Eq. (1) and its new parameter, the growth ratio, are definitely associated with the realm of *natural* processes. According to scientific methodological criteria of validation of scientific theories, finding such relationships should be considered as a one more argument in favor of validity of the growth equation and of the general growth mechanism.

### 4.4. *Amoeba's metabolic properties*

Once we know the total nutrient influx $K(t)$ and the growth ratio $G(t)$, we can find separately nutrient influxes for growth - $K_g(t)$, and maintenance -$K_m(t)$.

$$K_g(t)=\mathrm{K}(t)G(t) \tag{19}$$

$$K_m(t)=\mathrm{K}(t)\big(1-G(t)\big) \tag{20}$$

Also, we can find nutrient influxes per unit surface $k_s(t)=\mathrm{K}(t)/S(t)$ and per unit of volume $k_v(t)=\mathrm{K}(t)/V(t)$; accumulated amount of nutrients used for biomass synthesis $M_g$, maintenance $M_m$, and the total amount of consumed nutrients $M_{\mathrm{tot}}$ during the chosen time period [$t_1$, t]. We will use units of measure for the influx $\mathrm{pg}\cdot\mathrm{min}^{-1}$ ($1\mathrm{pg}=10^{-12}g$), except for *Amoeba*, for which the unit of measure is $\mu\mathrm{g}\cdot\mathrm{min}^{-1}$. Influx $k_s(t)$ is measured in $\mathrm{pg}\cdot\mathrm{min}^{-1}\cdot\mu\mathrm{m}^{-2}$ ; $k_v(t)$ in $\mathrm{pg}\cdot\mathrm{min}^{-1}\cdot\mu\mathrm{m}^{-3}$ (for *Amoeba*, accordingly $\mu\mathrm{g}\cdot\mathrm{min}^{-1}\cdot\mu\mathrm{m}^{-2}$ and $\mu\mathrm{g}\cdot\mathrm{min}^{-1}\cdot\mu\mathrm{m}^{-3}$).

$$M_g(t)=\int_{t_1}^{t}K_g(\tau)G(\tau)\mathrm{d}\tau \tag{21}$$

$$M_m(t)=\int_{t_1}^{t}K_m(\tau)G(\tau)\mathrm{d}\tau \tag{22}$$

$$M_{\mathrm{tot}}(t)=\int_{t_1}^{t}K(\tau)G(\tau)\mathrm{d}\tau \tag{23}$$

Application of Eqs. (19) - (23) to the growth curve in Fig. 1 produces metabolic characteristics presented in Fig. 4.

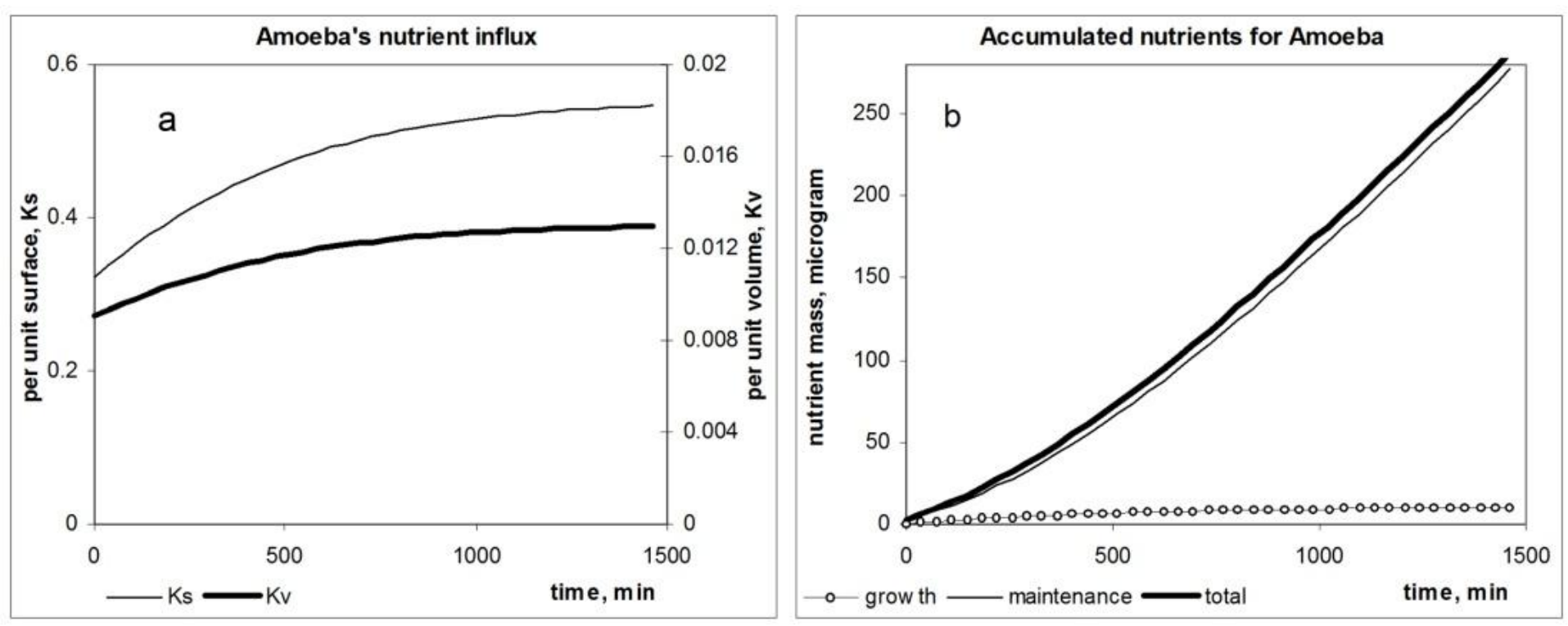

Fig. 4. Nutrient influx and accumulated amount of nutrients for *Amoeba*, depending on time. a - Specific nutrient influx $k_s(t)$ per unit of surface ( $\mu g \cdot min^{-1} \cdot \mu m^{-2}$ ) and per unit of volume $k_v(t)$ ($\mu g \cdot min^{-1} \cdot \mu m^{-3}$). b - Accumulated amount of nutrients used for growth and maintenance, and the total amount, in $\mu g$.

Metabolic properties of studied organisms will be compared in Table 1. For now, note that (a) *Amoeba* consumes about 28 times more nutrients for maintenance than for growth during the cell cycle; (b) we can find amount of synthesized biomass *directly*, while finding this critical for biotechnological applications parameter, in particular by methods of metabolic flux analysis, is a big problem today, in many respects.

### 4.5. *Amoeba's growth and division mechanism from the evolutionary perspective*

Geometrical form is an inherent property of *any* living organism and its constituents. Since the growth ratio is inherently tied to a geometrical form, the growth and division mechanisms based on direct changes of growth ratio are probably the most ancient ones. (We will call them as the growth and division mechanism of the *first type*.) Its characteristic features are as follows:

(1) The growth proceeds almost through the *entire* possible growth period (corresponding to the growth curve defined by the growth equation);

(2) The rates of protein and RNAs synthesis are the same;

(3) The value of the spare growth capacity is small, about 2%.

## 5. Fission yeast *S. pombe*. Growth and division

### 5.1. *Modeling growth of S. pombe using the growth equation*

This model organism represents the *second type* of a growth and division scenario. The computed growth curve in this case consists of convex and concave parts, separated by an inflection point. Such organisms do not go through the whole possible growth cycle, but use only the fastest, convex, part of the whole growth curve, switching to division much earlier, at the inflection point of the growth curve. This evolutionary enhancement secures the fastest possible growth time. Evolutionarily, such mechanism was very likely developed on top of more basic mechanisms, like the ones studied in *Amoeba*, since it requires a set of advanced features, which unlikely appeared simultaneously from scratch.

For illustration, we used experimental data, courtesy of Baumgartner and Tolic-Norrelykke[45]. Earlier, in Ref. 22, similar results were obtained for 85 experiments from the same study, for the temperatures of $32\ ^0C$, $28\ ^0C$ and$25\ ^0C$, and also for experimental graphs from Ref. 46. Therefore, the presented results can be considered as statistically meaningful.

We use the same geometrical model of *S. pombe* as in Refs. 20,22. The organism is modeled by a cylinder with a length *l*, radius *r*, with hemispheres at the ends; beginning length of a cylinder is $l_b$, its ending length is $l_e$. In these notations, using Eqs. (2) - (4), the relative surface, relative volume and the growth ratio can be found as follows.

$$R_{Sc} = \frac{(2r+l)}{(2r+l_e)};\ R_{Vc} = \frac{((4/3)r+l)}{((4/3)r+l_e)}; \qquad (24)$$

$$G_c = \frac{R_{Sc}}{R_{Vc}} - 1 = \frac{((4/3)r+l_e)(2r+l)}{((4/3)r+l)(2r+l_e)} - 1 \qquad (25)$$

where index '*c*' denotes 'cylinder'. (Formulas above are easier understood if one takes into account that two hemispheres at the ends constitute a sphere.)

We will also need the relative lengths' increases $L = l/l_b$ and $E = l_e/l_b$, and a relative to $l_b$ radius's increase $R = r/l_b$. Then, the growth ratio from Eq. (25) can be rewritten as follows.

$$G_c = \frac{(2/3)R(E-L)}{((4/3)R+L)(2R+E)} \qquad (26)$$

(Equation (26) is obtained from (25) by transformation, and then by division of a numerator and denominator by $l_b^2$.) Volume *V* of a cylinder with hemispheres is $V = (4/3)\pi r^3 + \pi r^2 l$. The differential is $dV = \pi r^2 dl = \pi r^2 l_b d(l/l_b) = \pi r^2 l_b dL$. The nutrient influx is defined by Eq. (7).

Note that Eq. (26) uses the relative length's increase for the *cylindrical* part of the organism, not for the *whole* length. The rationale is that the cell's volume increases through the elongation of the *cylindrical* part. (This consideration is not critical - using the whole length produces close results.)

A diameter of a cylinder does not change during the growth. Of course, the growth equation allows considering scenarios with a changing diameter too, and the appropriate growth curves were calculated as well. However, in this case, we cannot obtain an analytical solution, and have to use a numerical one.

Substituting these parameters into Eq. (1), we obtain the following differential equation.

$$p\pi r^2 l_b dL = A(C_p L^2 + C_r L^3)\frac{(2/3)R(E-L)}{((4/3)R+L)(2R+E)}dt$$

Separating variables and doing appropriate transformations, we obtain the following.

$$\left(\frac{3p\pi R l_b^3 (2R+E)}{2C_p}\right)\frac{(4/3)R+L}{L^2(1+(C_r/C_p)L)(E-L)}dL = Adt \qquad (27)$$

The analytical solution of this equation is as follows[20,22]. Let us denote $C = C_r / C_p$ and $B = \frac{3p\pi R l_b^3 (2R+E)}{2C_p}$, which are constants. For simplicity, we first find the integral

$$\int_{L_b}^{L} \frac{(4/3)R + L}{L^2(1+CL)(E-L)} dL$$

and then multiply the found solution by $B$.

We find the integral of the left part as a sum of integrals. In order to do that, we use the presentation with unknown values of $d, f, g, h$ to be found.

$$\frac{(4/3)R + L}{L^2(1+CL)(E-L)} = \frac{Ld + f}{L^2} + \frac{g}{1+CL} + \frac{h}{E-L} \quad (28)$$

Transforming the right part of this equation to a common denominator, we obtain the following.

$$\frac{L^3(hC - g - dC) + L^2(dCE - d - fC + gE + h) + L(dE - f + fCE) + fE}{L^2(1+CL)(E-L)}$$

Equating coefficients for the appropriate powers of $L$ in this equation, and the right part of equation (27), we obtain the following system of four equations with four unknown values.

$$hC - g - dC = 0$$
$$dCE - d - fC + gE + h = 0$$
$$dE - f + fCE = 1$$
$$fE = (4/3)R$$

Solving this system, we find.

$$f = \frac{4R}{3E};\ d = \frac{1 + f - fCE}{E};\ h = \frac{d + fC}{CE + 1};\ g = C(h - d)$$

Substituting the right part of Eq. (28) into differential equation (27), and doing integration, we finally obtain

$$t = \frac{B}{A}\left( f\left(\frac{1}{L_b} - \frac{1}{L}\right) + d\ln\left(\frac{L}{L_b}\right) + \frac{g}{C}\ln\left(\frac{1+CL}{1+CL_b}\right) + h\ln\left(\frac{E - L_b}{E - L}\right)\right) \quad (29)$$

Although we would prefer to have a functional dependence of $L = L(t)$, the transcendental Eq. (29) has no analytical solution. However, it is still much more convenient to have Eq. (29) than doing numerical integration, since it allows easily obtaining two sets of values for time and length, and then one can choose time as an independent argument. Note that the last term in Eq. (29) provides a vertical asymptote, that is when $L \to E$, $t \to \infty$. Accordingly, the reverse function $L(t)$, which, from a geometrical perspective, is a symmetrical reflection of function $t(L)$, defined by Eq. (29), relative to the line $L = t$, has a term with a horizontal asymptote.

The parameter $L_b$ in (29), the relative beginning length, is shown rather for consistency. In our notations $L_b = 1$.

Unlike in *Amoeba*, the rate of RNA synthesis in *S. pombe* is about double of the rate of protein production[20,22]. This is why we obtained the cube of length in (7).

It is often assumed[45] that the double rate of RNA synthesis triggers *after* completing S phase, while before that the rates of protein and RNA synthesis are the same. In this case, $K(L)=AL^2(C_p+C_r)$, and the solution of the growth equation is as follows[22].

$$t=\frac{B_S}{A_S}\left(f_S\left(\frac{1}{L_b}-\frac{1}{L}\right)+d_S\ln\left(\frac{L}{L_b}\right)+g_S\ln\left(\frac{E-L_b}{E-L}\right)\right) \qquad (30)$$

where $B_S=3p\pi Rl_b^3(2R+E)/(2(C_p+C_r))$; $f_S=4R/(3E)$; $g_S=d_S=(1+f)/E$.

Model's input parameters are listed in Table 1. A diameter and a fraction of nutrients used for RNA synthesis were estimated based on the fact of fast growth and analogy with other microorganisms, like in Ref. 36, and large initial size of the considered species. Unfortunately, these parameters were not measured by researchers.

As we can see from Fig. 5a, *S. pombe*, unlike *Amoeba*, does not proceed through the whole possible growth cycle, defined by the full growth curve, but switches to division phase *at inflection point*, which secures the minimum growth time at a maximal possible rate of biomass production (this can be proved mathematically). This significant evolutionary enhancement secures much faster growth with relatively less nutrients used for maintenance than in case of *Amoeba*, so that more nutrients are used for growth.

The value of the spare growth capacity (SGC) for *S. pombe* is much greater than *Amoeba*'s 2%, and resides in the range of 30-40%. However, knowledge of SGC for computing growth curves in case of *S. pombe* and similarly growing organisms (including *B. subtilis, E. coli*) is not required, since comparison with experimental data is based on the beginning of division, which coincides with the inflection point.

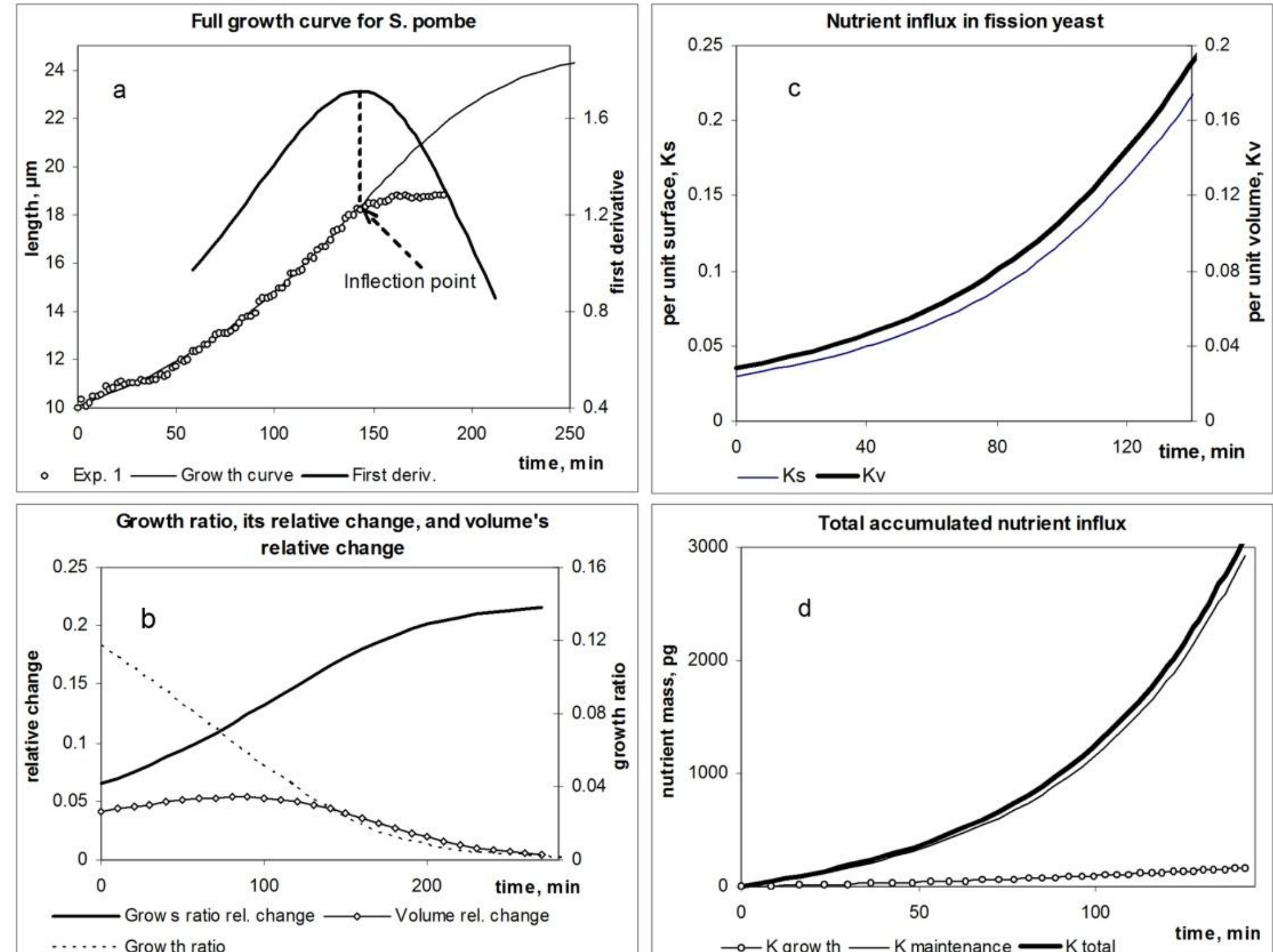

Fig. 5. *S. pombe's* growth and metabolic characteristics. a - Full growth curve for *S. pombe* versus experiment 1 from $32\ ^0C$ dataset from Ref. 45. Maximum of the first derivative of the growth curve corresponds to the beginning of division phase and inflection point of the growth curve. b - Change of the growth ratio, and the relative changes of the growth ratio versus the relative change of volume, for equal time intervals. c - Nutrient influx per unit surface $k_S$ (measured in $pg \cdot \mu m^{-2} \cdot min^{-1}$) and per unit of volume $k_V$ (measured in $pg \cdot \mu m^{-3} \cdot min^{-1}$). d - Accumulated nutrients for maintenance, growth and the total amount, in pg.

As it was the case with *Amoeba*, the relative (to the total nutrient influx) amount of produced biomass, defined by the growth ratio, remains the important parameter, which defines composition of biochemical reactions through the growth cycle. However, in *S. pombe*, it triggers the beginning of a division phase at the inflection point of the growth curve.

High value of SGC (and accordingly the possibility of continuing to grow beyond the inflection point) for *S. pombe* is not a mathematical extravagance, but a *real* property of such cells. Indeed, cells can grow substantially bigger than their normal size, when the division is suppressed[7]. For *S. pombe*, it was confirmed experimentally in Ref. 45. Computations in Ref. 22, on the basis of Eq. (30), confirmed this as well, and produced a growth curve similar to experimental data.

Metabolic properties of *S. pombe* were studied using Eqs. (19) - (23). Fig. 5b is presenting further evidence that the relative amount of produced biomass is that parameter which drives growth and division process of *S. pombe*, forcing changes in the composition of biochemical reactions in such a way that the organism proceeds through its life cycle. Indeed, we can see that the relative change of the growth ratio computed at equal time intervals is *substantially greater* than the relative change of volume (by 3.8 times at the division point). Also, the rate of change of the growth ratio quickly *increases* at the beginning of the division phase, while the relative change of volume *decreases*. So, it is very unlikely that changes in volume (or of any absolute dimensional parameter) could be a factor triggering *S. pombe's* division, besides the fact that changes in volume do not explain, why the same species, which could differ in size at the division point as much as four times, divide; or why such organisms continue to grow once the division is suppressed, and so on. The general growth mechanism, in this regard, explains *all* known properties and facts about growth and division of such organisms, explains why they have certain shape, allows finding their metabolic properties, and so on.

Fig. 5c shows change of specific nutrient influxes $k_s(t)$ and $k_v(t)$. Unlike in *Amoeba* (see Fig. 4a), the increase of these influxes accelerates all the time. This is also a factor contributing to fast reproduction. Fig. 5d shows amount of accumulated nutrients for growth and maintenance, and the total amount of consumed nutrients. Note that maintenance of *S. pombe* requires about 18.2 times more nutrients than biomass production, while in *Amoeba* this ratio was equal to 28. This is a definite and significant evolutionary improvement optimizing the use of nutrients for fast reproduction.

The obtained results also address a long debated issue, is *S. pombe's* growth curve exponential or piecewise linear. Eqs. (28) and (30) answer the question - neither one in a pure form. However, given the presence of logarithmic functions, the reverse dependencies (producing the growth curves in question) are rather closer to exponential functions than to piecewise linear dependencies. Ref. 22 presents statistical evidence in this regard.

### 5.2. *Growth and division mechanism of the second type*

The considered second type of growth and the division mechanism very much differ from the same mechanisms of the first type, used by *Amoeba*. (Note that both types of growth also have very different characteristics of population growth[47] too, with the second type to be much more resilient to extinction.)

The following features are characteristic for the second type of growth:

(1) The growth curve has a well expressed inflection point;

(2) Species do not go through the entire possible growth cycle, but switch to division much earlier, at the inflection point of the growth curve, thus minimizing the growth time, while maximizing the use of nutrients for growth. At inflection point, the rate of change of produced biomass is the highest, which also means the highest rate of change of composition of biochemical reactions. These two factors secure reliable triggering of division phase;

(3) Such species are elongated (the inflection point is better expressed for elongated forms);

(4) The rate of RNA synthesis is double the rate of protein synthesis (which is also a factor contributing to better expression of an inflection point and faster growth);
(5) If the division is suppressed in experiments, such cells continue to grow further, in accordance with the theoretically calculated growth curve. This occurs because of the high value of SGC, as it was revealed by the growth equation. SGC represents a qualitative measure of unrealized growth potential for such organisms.

## 6. Dependence of growth rate on geometrical form

Note that according to the general growth mechanism, among all elongated forms a cylinder has the fastest growth time due to a higher value of the growth ratio[20,22]. *This* is why many microorganisms, both sedentary and actively moving, have a cylinder shape. Fig. 6 shows such dependence graphically for a double frustum whose base changes from zero to a base's diameter, that is from a double cone to a cylinder. Here, we assume that nutrient influx per unit surface is the same and constant for all forms.

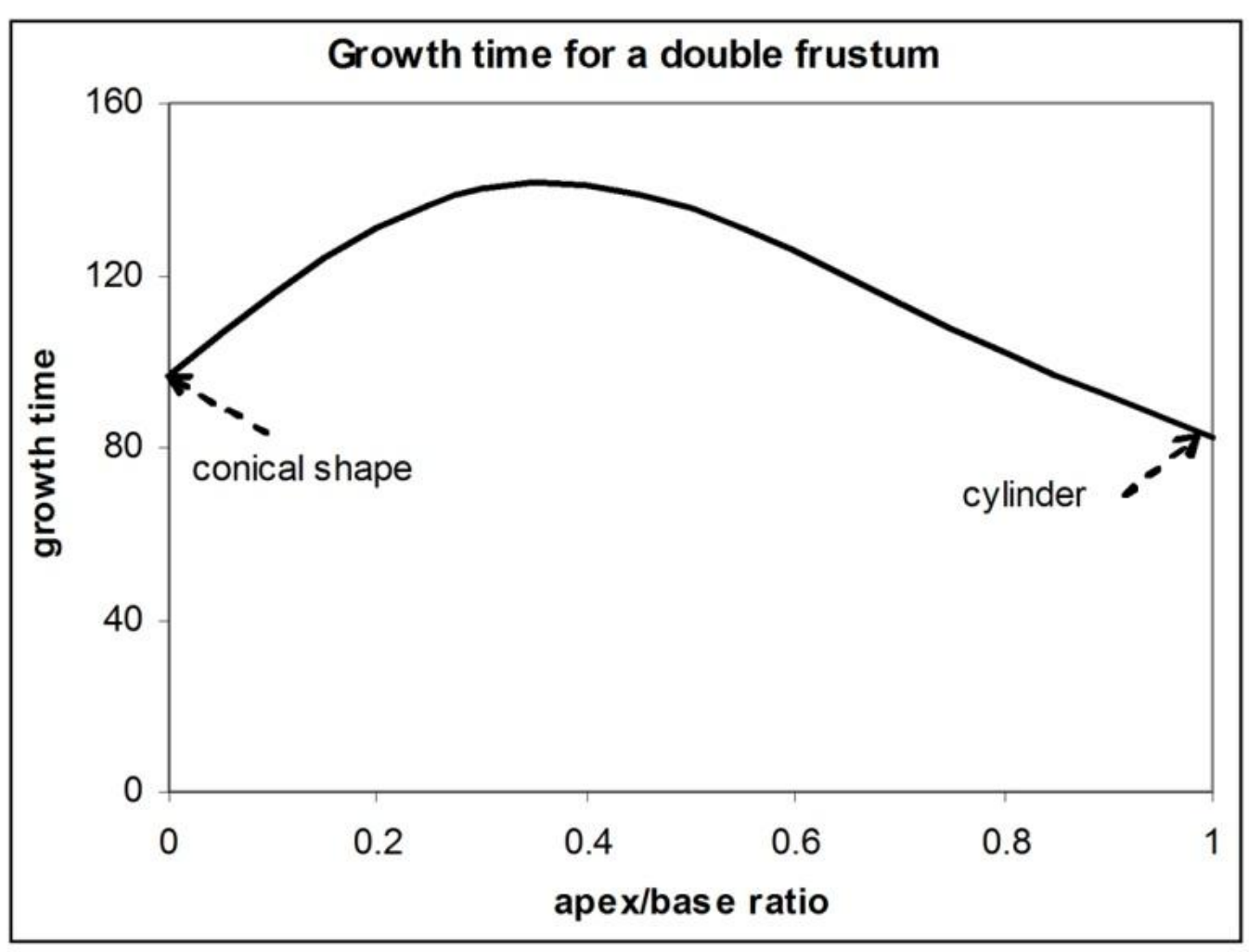

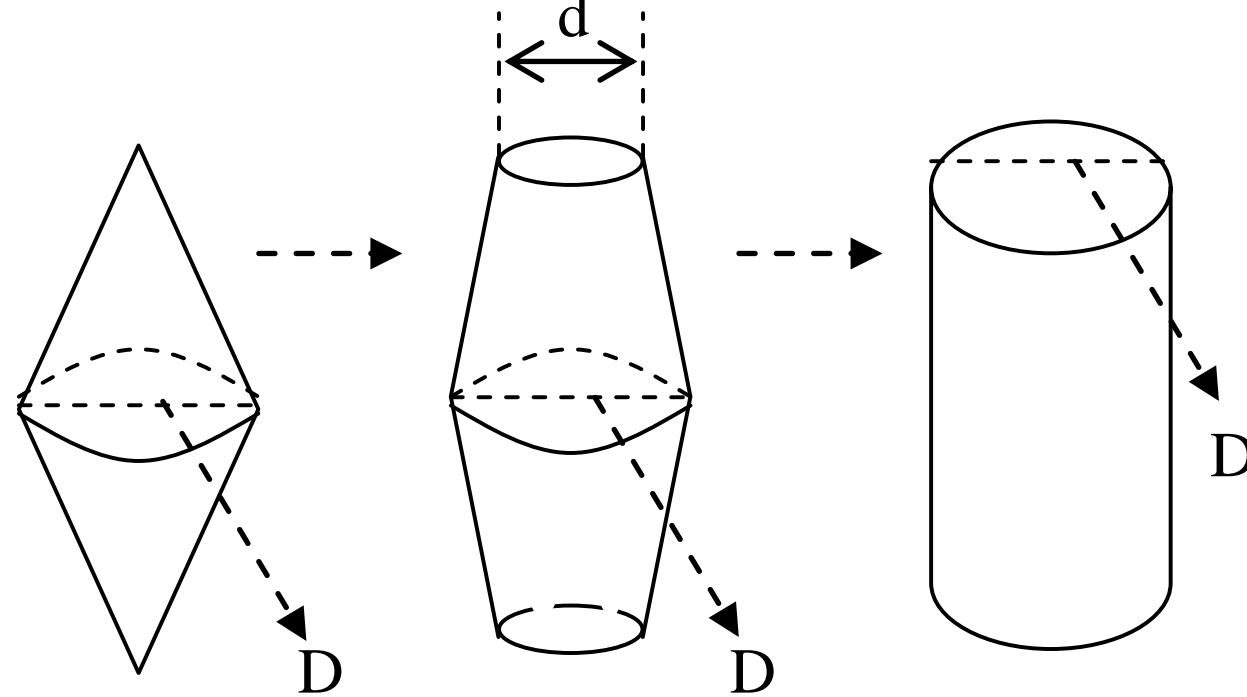

Fig. 6. Growth time for a double frustum depending on the relative upper apex's diameter $d/D$ (upper panel). Change of the upper apex diameter $d$ from zero to $D = 1$ (bottom panel) provides transition from a double cone shape (leftmost point) to a cylinder (rightmost point).

The shortest growth time corresponds to a cylinder. The second fastest growing form is a double cone. On the other hand, a certain shape of a double frustum has the slowest growth time (the point of maximum of the curve in Fig. 6); in other words, such a form *suppresses* the growth. And, indeed, this effect, that is suppression and significant growth deceleration by changing the cells' geometry, for example from spherical to ellipsoidal, is realized in nature. Note that this effect was first discovered *theoretically* by the author, and *then* experimental confirmations were found[43,44], as it was earlier mentioned.

Maximizing the growth ratio (in other words, maximizing the amount of produced biomass) is one of the reasons why so many elongated microorganisms have a cylinder shape. The argument that such a shape is due to the need for a lesser resistance during motion does not sustain - many immobile or low motility organisms also have a cylinder form, like *B. subtilis*. Also, the cylinder form is less restrictive with regard to the maximum length, since the value of the growth ratio changes slowly towards the end of growth for elongated forms. (This effect and the earlier triggering of division, in turn, explain large variations in the relative length's increase of *S. pombe* and other elongated organisms.)

Overall, all known facts about growth of *S. pombe*, overgrowth and its cylindrical form, as well as about similar characteristics of other elongated organisms and cells, are well explained by the general growth mechanism.

A side note. Fig. 6 shows that the second elongated fastest growing form is a double cone. Shape of a carrot could be a consequence of this effect. Even though the nutrient and water supply go through a network of interacting xylem and phloem flows, in case of a carrot, and plants in general, deposition of sugars into sink cells is done through the *surface*, although this is an *internal* surface this time (unpublished study). Thus, the evolutionary growth optimization could have this fact as a reason, which led to a carrot's cone shape.

A sphere is also a fast growing form. Although its growth curve does not have a well expressed inflection point, and such organisms should not enter division prematurely, as *S. pombe* does, the sphere's growth ratio at the beginning is more than two times greater than that of a cylinder with the ratio of length to a diameter of 2 : 1. This is why many unicellular organisms have a spherical shape, as well as many fruits and vegetables, especially when the vegetation period is short, which is the case for Northern berries, apples. Overall, geometry and interaction of relative surface and relative volume, reflected in the value of the growth ratio, governs the growth of plants too, although in a more complicated, transformed form, influenced by other factors and adaptation mechanisms to specific environments.

## 7. Growth and division of *B. subtilis*, *E. coli*

*S. pombe, B. subtilis* and *E. coli* exercise the growth and division mechanisms of the second type, although the first one is eukaryote, the other two are bacteria. Nutrient influx for them is defined by Eq. (7), since, as it was previously discussed, *E. coli* has a double rate of RNA synthesis compared to protein synthesis. There are no such data for *B. subtilis*, but it can be assumed the same, with very high probability, given the similarity of geometrical forms of *E. coli* and *B. subtilis* and their fast growth. Both factors, according to the results for *S. pombe*, strongly correlate with a double rate of RNA synthesis.

Fig. 7a,b show computed growth curves for *E. coli* (by Eqs. (29)) versus two experimental data sets from Ref. 48. Fig. 7c shows a similar growth curve for *B. subtilis* versus experimental data from Ref. 49. Note, the data fit, used in Ref. 49, belongs to authors of Ref. 49, which used multiple measurements for that, but this is *not our* data fit. We use the same approach as it was declared at the beginning, that is we compute growth curves using *only* the growth equation with strictly necessary input parameters, and only *after that* compare the computed growth curves with experimental data. In all instances, we see a *very good* correspondence between the *computed growth curves* and experimental dependencies. Model's input parameters are listed in Table 1. The fraction of nutrients for RNA synthesis was estimated based on the rate of growth (the higher the rate of growth is, the greater this fraction) and the possible range of this value (0.035 to 0.246 for *E. coli*, according to Ref. 36.) The diameter was estimated based on geometrical proportions of organisms.

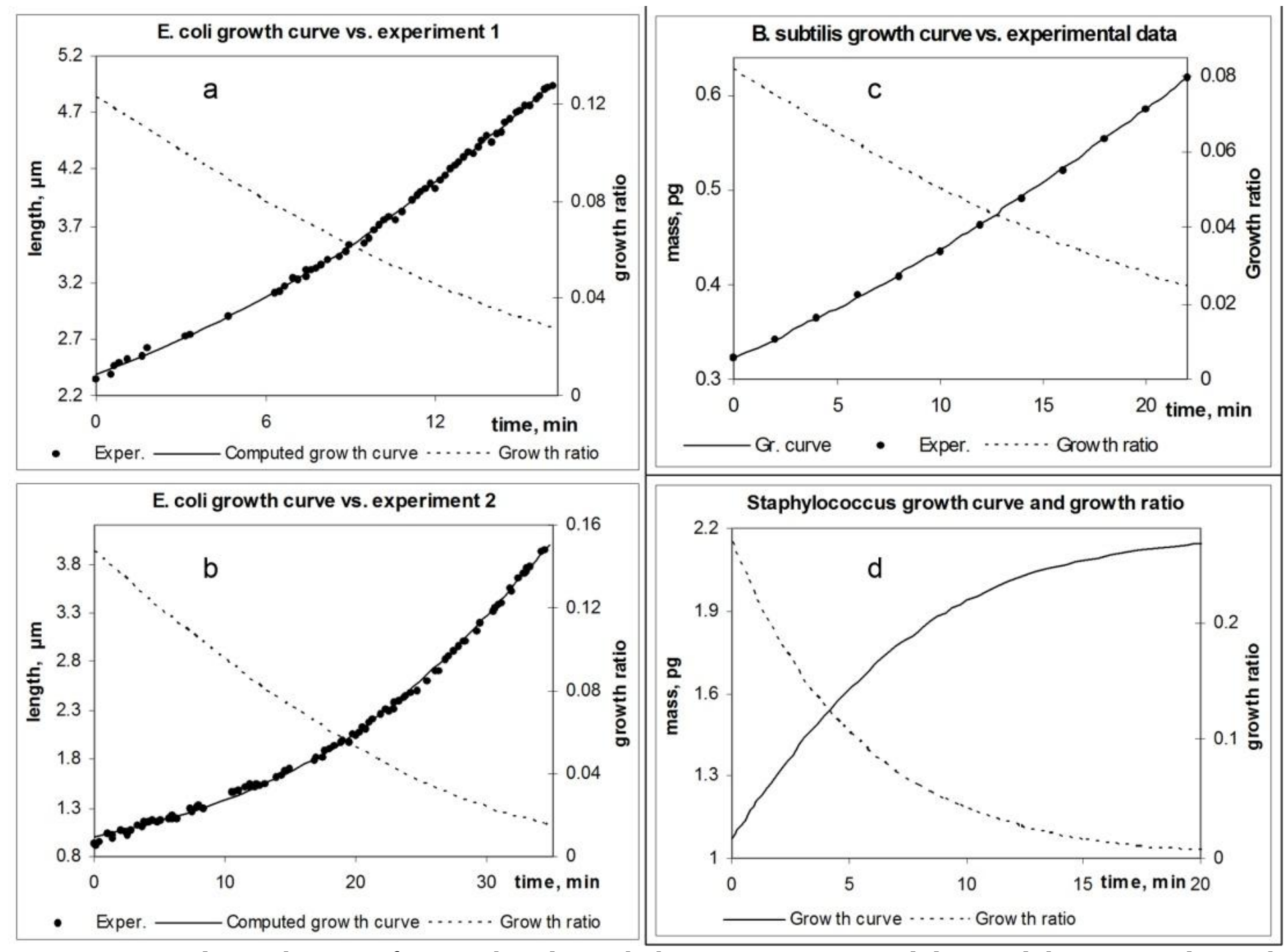

Fig. 7. Computed growth curves for *E. coli* and *B. subtilis* versus experimental data, and the computed growth curve for *Staphylococcus*. Experimental data for *E. coli* are from Ref. 48; for *B. subtilis* - from Ref. 49. a - The growth curve for *E. coli* vs. the data the authors suggested to model by a bilinear curve. b - The same for the data the authors suggested to model by a tri-linear curve. c - a computed growth curve for *B. subtilis* versus experimental data. d - a computed growth curve and the growth ratio for *Staphylococcus*.

It was suggested in Ref. 48 approximating *E. coli's* growth curve as a bilinear or tri-linear function. If we take into account rounding of the tip of a divided microbe in the first minutes of growth, which is the cause of faster length's increase at the very beginning, then, the computed growth curves actually correspond to experiments noticeably better than the authors' bi- and tri-linear approximations. The length's increase due to the tip rounding at the beginning of growth was proved in Ref. 45, and later was confirmed in Refs. 21, 22.

For *E. coli* and *B. subtilis*, metabolic properties (the graphs for nutrient influx and accumulated nutrients for growth and maintenance) are similar to the ones for *S. pombe*, that is they are quickly increasing convex curves. For *Staphylococcus*, the appropriate metabolic curves resemble the concave curves for *Amoeba*.

## 8. Growth and division model of *Staphylococcus*

*Staphylococcus's* growth was modeled by an increasing sphere. It turned out to be an interesting model from many perspectives, including proofs of validity of the growth equation. The rate of RNA and protein synthesis was assumed the same. (Although, there is a possibility that the rate of RNA synthesis can be greater, similar, for instance, to *S. cerevisiae*[37].) Then, Eq. (9) transforms into

$$K(v)_s = A_s(C_p + C_r)v^{4/3} \tag{31}$$

(index '*s*' denotes 'sphere').

The growth ratio can be found as follows.

$$R_{Ss} = \frac{R^2}{R_0^2};\ R_{Vs} = \frac{R^3}{R_0^3};\ G_s = \frac{R_S}{R_V} - 1 = \frac{R_0}{R} - 1 \tag{32}$$

where $R_0$ is the maximum possible (asymptotic) radius; *R* is the current radius.

Substituting $G_s$ from (32), and nutrient influx from (31) into Eq. (1), and remembering that we substitute the term *kS* in (1) by the term $K(v)_s$, similar to what we did to obtain Eq. (13), we obtain the following differential equation.

$$4\pi p R^2 dR = A_s(C_p + C_r)\left(\frac{R}{R_b}\right)^4 \left(\frac{R_0}{R} - 1\right) d\tau$$

which can be presented as

$$d\tau = \frac{4\pi p R_b^4}{A_s(C_p + C_r)(R_0 - R)R} dR \tag{33}$$

Mathematically, the solution of (33) is very similar to solution of (13) for *Amoeba*, and it is as follows.

$$t = \frac{4\pi p R_b^4}{A_s(C_p + C_r)R_0} \ln\left(\frac{r(R_0 - R_b)}{R_b(R_0 - r)}\right) \tag{34}$$

where the new variable *r* denotes the radius of a sphere at time *t*.

As we can see, it is very similar to solution (14) of the growth equation for *Amoeba* growth. Denoting

$$c_{0s} = \frac{4\pi p R_b^4}{A_s(C_p + C_r)R_0}$$

and solving (34) for radius *r*, we obtain a solution

$$r = \frac{R_0 \exp(t / c_{0s})}{(R_0 / R_b - 1) + \exp(t / c_{0s})},$$

which is *exactly* the same as Eq. (15) for *Amoeba*, save for the constant term $c_{0s}$. In other words, we *again* obtained a generalized solution of a logistic equation, same as for *Amoeba*! Certainly, thus obtained equation has the same interesting properties - close relationship with a logistic equation, and the exponential dependence of the growth ratio on time. So, for now, we found two geometrical forms, a disk and a sphere, possessing these two interesting properties. Both from the mathematical and physical perspectives, these results are of great interest and importance, representing some fundamental properties of the real world, including its geometry, which so far were unknown. These and previous results allow to conclude that the growth equation represents a *new equation of mathematical physics*, whose study could provide new important insights to the properties of our world.

Division mechanism of *Staphylococcus* is very likely of the first type, as in *Amoeba*. The reasons for such a suggestion is that the growth curve of a sphere does not have a well expressed inflection point, or other specific features, which could serve as the checkpoints for starting an earlier division without going through the whole possible growth curve. We can see from Fig. 7d that the growth ratio is much higher in *Staphylococcus* at the beginning of growth (a value of 0.27 versus 0.117 for *S. pombe*). This accordingly means a 2.3 times greater fraction of nutrients going to biomass production at the beginning, which apparently compensates for slower growth at the end.

Thus, we made a set of predictions regarding growth and division mechanisms of *Staphylococcus* and other organisms having a spherical shape, based on the general growth mechanism. These predictions certainly can, and in the author's view should, be tested experimentally. Certainly, new effects and specifics will be discovered during such experiments, since Nature is inexhaustible for new things, but the main growth and division properties found for *Staphylococcus* should stay. An open question though remains about the rate of RNA synthesis. If it is greater than that of protein, then the growth curves will differ. An experiment could introduce clarity in this regard. As it well known, theory is one thing, the practice is another venue.

## 9. Metabolic properties of cells. Allometric scaling

Table 1 presents the summary of metabolic properties of considered organisms and input parameters for their models. Found metabolic characteristics can serve as an additional verification of obtained results and of the general growth mechanism and growth equation. If they are correct, then

(a) we should obtain the allometric scaling coefficient within the experimentally found range;

(b) the dependence of metabolic rates on volume, presented in logarithmic coordinates, should be close to a linear one.

In addition to the earlier introduced characteristics, the following metabolic parameters were calculated: average and maximal metabolic rates per unit surface $k_{\mathrm{Sav}}$ and $k_{S\mathrm{max}}$; average and maximal metabolic rates per unit volume $k_{\mathrm{Vav}}$ and $k_{V\mathrm{max}}$; the total maximal, average and minimal nutrient influxes, accordingly $K_{\mathrm{max}}$, $K_{\mathrm{av}}$, $K_{\mathrm{min}}$.

Table 1 considers nutrient influxes, but not the actual metabolic rates. We assume that the amount of produced energy is proportional to consumed nutrients[34]. On one hand, using nutrients has an advantage over the conventional methods, which may not account for all metabolic mechanisms. In particular: (a) it becomes possible to compare the consumed amount of food with the measured metabolic output; (b) knowing metabolic mechanisms of particular organisms, it is possible to translate the amount of consumed nutrients into the metabolic output. On the other hand, metabolic output for the same amount of nutrients can differ in different organisms. Besides, different types of nutrients could provide different metabolic outputs. However, the influence of these factors, if they indeed exist, could be only by an order of magnitude less.

Note the large variations of nutrient influxes *per unit volume* (up to 80 times) between different organisms in Table 1, while the nutrient influx *per unit surface* differs little (of about *4 times*). If we think for a moment, this is understandable, since nutrients are acquired by considered cells through the *surface*, from the common nutritional environment, and so the differences, indeed, should not be as great, being much dependent on the concentration of nutrients in the surrounding environment. It is interesting that from this observation a well founded theory of interspecific metabolic allometric scaling in unicellular organisms was developed, which exposed fundamental causes of this phenomenon and led to far reaching conclusions[34]. This can serve as a one more example of usefulness of the general growth mechanism for diverse biological studies.

We can verify the validity of obtained in Table 1 metabolic parameters using a metabolic allometric scaling effect[34]. Fig. 8a presents such a dependence[34]. The found values of metabolic rate, indeed, are located on a straight line, and the value of allometric exponent of 0.758 complies with results from Ref. 50, according to which allometric exponents for unicellular organisms are in the range from 2/3 to more than one. So, the verification condition (a), we set at the beginning, is fulfilled. Fig. 8a shows that the condition (b) is also fulfilled – the dependence of metabolic rate on volume, in a logarithmic scale, is indeed a linear one, as it should be, if the general growth mechanism and the growth equation are correct. Thus, we obtained one more proof of their validity. (With the assumption that the food chain for microorganisms has similar properties as for multicellular organisms; this consideration also has experimental confirmations).

The obtained in Table 1 results are also very important from another perspective. Using this result, it was shown in Ref. 34 that the entire food chain is organized in such a way that it preserves its continuity, on one hand, and a dynamic balance between different species composing the food chain, so that none of them could have an overwhelming advantage, on the other hand. This important result is due to the general growth mechanism and growth equation as well. So, we found an additional proof of validity of the general growth mechanism, the growth equation, and of their high value for studying diverse problems in biology and related disciplines.

However, the results in Table 1 and Fig. 8b are even more revealing. Fig. 8b presents dependence of nutrient influx *per unit surface* for different unicellular organisms depending on mass, in logarithmic scale. We can see that except for *E. coli*, which is a *highly motile* organism, the values of nutrient influx for other microorganisms, which are all *sedentary*, very well fit a straight line. What does it mean? It means that sedentary organisms exploit the nutritional environment, meaning the amount of taken nutrients per unit surface, according to strict hierarchy defined by their mass. This fact also has important implications, resulted in discovery of an evolutionary mechanism of food chain creation, as well as obtaining other important results and explaining prior known but not understood facts and effects[34]. Together, these new findings also provide additional proofs of validity of the general growth mechanism and of the growth equation.

If we look at this point back, we can see how far we went: from the introduction of the growth equation up to metabolic allometric scaling properties of unicellular organisms and fundamental properties of food chains! And, at each of such landing spots on our way, without a single exception(!), we found solid proofs of validity of results, obtained through the use of growth equation and concepts of the general growth mechanism. These results make a strong case in favor of the thesis which was set forth at the beginning: Life phenomena are governed by a tight cooperative working of physical and biochemical mechanisms, but not by biochemical mechanisms alone, as the present biochemical paradigm of Life assumes.

Table 1. Summary of metabolic properties of considered organisms and input parameters for their growth models.

| | B. subtilis | Staphylo coccus | E. coli, 3-Linear | E. coli, 2-Linear | S. pombe | *Amoeba* |
|---|---|---|---|---|---|---|
| $k_{Sav}$, $pg\cdot\mu m^{-2}\cdot min^{-1}$ | 0.083 | 0.135 | 0.1053 | 0.2194 | 0.0957 | 0.493 |
| $k_{S\,Max}$, $pg\cdot\mu m^{-2}\cdot min^{-1}$ | 0.134 | 0.147 | 0.284 | 0.433 | 0.222 | 0.545 |
| $k_{Vav}$, $pg\cdot\mu m^{-3}\cdot min^{-1}$ | 0.672 | 0.5336 | 1.0073 | 0.8933 | 0.0866 | 0.01215 |
| $k_{Vmax}$, $pg\cdot\mu m^{-3}\cdot min^{-1}$ | 1.066 | 0.552 | 2.65 | 1.71 | 0.195 | 0.013 |
| $V_{max}$, $\mu m^{-3}$ | 0.617 | 2.145 | 3.999 | 16.975 | 325.4 | 1.88E+7 |
| Diameter, $\mu m$ | 0.536 | | 0.446 | 1.09 | 5 | |
| Diameter beg. $\mu m$ | | 1.27 | | | | 285.7 |
| Diameter end, $\mu m$ | | 1.6 | | | | 409.5 |
| Beginning length, $\mu m$ | 1.608 | | 1.003 | 2.39 | 10.1 | |
| Ending length, $\mu m$ | 2.915 | | 3.999 | 4.957 | 18.24 | |
| Asymptotic length, $\mu m$ | 4.7 | | 6.563 | 7.641 | 24.9 | |
| $k_{av}$ , $pg\cdot min^{-1}$ | 0.3538 | 0.98 | 0.26 | 2.86 | 4.984 | 1.96E+5 |
| $K_{max}$, $pg\cdot min^{-1}$ | 0.658 | 1.184 | 1.59 | 7.34 | 63.59 | 2.44E+5 |
| $K_{min}$, $pg\cdot min^{-1}$ | 0.116 | 0.47 | 0.025 | 0.688 | 4.723 | 8.28E+4 |
| $M_g$ (nutr. growth), $pg$ | 0.295 | 1.106 | 0.472 | 2.41 | 160.74 | 9.95E+6 |
| $M_m$ (nutr. maint.), $pg$ | 6.82 | 18.49 | 13.33 | 43.64 | 2930 | 2.77E+8 |
| $M_t$ (nutr. total), $pg$ | 7.12 | 19.6 | 13.8 | 46.05 | 3091 | 2.87E+8 |
| $M_m / M_g$ | 23.1 | 16.72 | 28.2 | 18.12 | 18.23 | 27.83 |
| Fraction of nutrients for RNA synthesis, % | 12 | | 6 | 10 | 60 | |
| Logarithm $k_{Vav}$ | -0.397 | -0.628 | 0.0073 | -0.113 | -2.447 | -4.41 |
| Logarithm $k_{Vmax}$ | 0.064 | -0.59 | 0.97 | 0.538 | -1.64 | -4.35 |
| Logarithm $V_{max}$ | -0.48 | 0.76 | 1.386 | 2.832 | 5.785 | 16.75 |
| Logarithm $K_{max}$ | -0.42 | 0.169 | 0.465 | 1.994 | 4.152 | 12.405 |

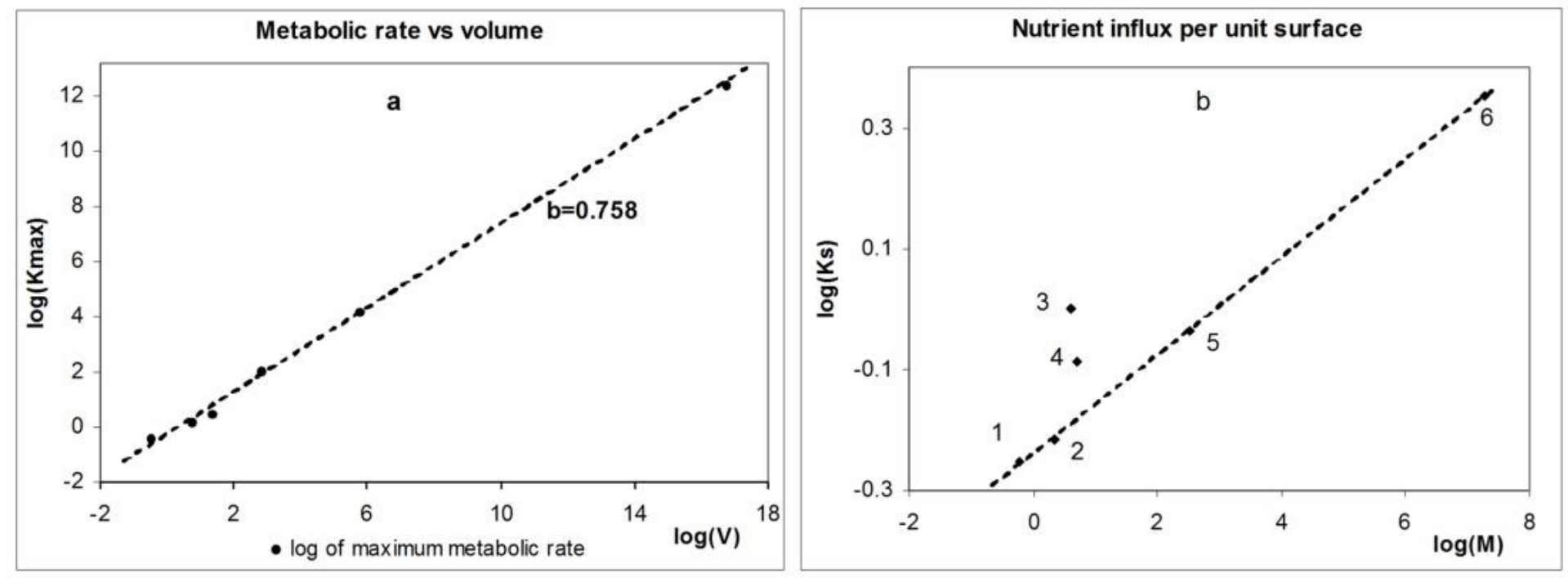

Fig. 8. Change of metabolic rate and nutrient influx depending on volume of organisms, in logarithmic scale. Numbered data points from left to right correspond to *B. subtilis* (1), *Staphylococcus* (2), *E. coli 1* (3), *E. coli 2* (4), *S. pombe* (5), *Amoeba* (6). a - maximal total metabolic rate. b - metabolic influx per unit surface. (Graphs are from Ref. 34.)

## 10. Recovery of growth rate in unicellular organisms with damaged biochemical machinery as a proof of validity of the general growth mechanism

Suppose we have organisms with damaged biochemical machinery, including deletion of genes, whose rate of biomass synthesis (or the rate of growth) was accordingly significantly reduced because of the inflicted damage. In Ref. 52, studying *E. coli*, such scenario is defined as "*a situation that cripples growth (growth rate, 20% of wild-type levels in glucose minimal media ...) but does not kill the organism*." Then, if the general growth mechanism indeed exists, then its action should be exhibited as a tendency of such organisms to gradually, through successive generations, recover the growth rate to the value defined by the growth ratio. If such a mechanism is a fiction, then such organisms would rather remain in a crippled state, or will experience random biochemical changes, which, besides other developments, also might improve the rate of growth. However, such improvement would be *definitely* different for different organisms in this case. Indeed, for such improvement to be the same (in terms of some universal invariant) for all different damaged organisms, such invariant must be defined *above* molecular level, because compositions of biochemical reactions in different organisms, and even in their different strains, are *different* after recovery, exercising different pathways (as we will see soon based on the available literature).

### 10.1. *Diversity of recovered pathways*

Note that although for brevity we cite here only few papers in this regard, the literature on the subject is extensive. In Ref. 53, studying *E. coli*, the authors acknowledge:
"*These results demonstrate that a functional pathway can be patched together from promiscuous enzymes in the proteome, even without mutations in the genes encoding those enzymes*."

"*Evolution of novel pathways continues in the present "Promiscuous" activities - which we define as adventitious secondary activities that are physiologically irrelevant - are extremely common.*"

"*Consequently, if the environment changes and a promiscuous activity becomes important for fitness, it can serve as a starting point for evolution of a new enzyme. The potential for assembly of novel pathways using the resources in a given proteome is unknown, but likely considerable*."

"*there could be 16,000 potential reactions whose circuitry could be rewired to generate novel pathways*."

"*We were able to evolve a novel 4-step SP in E. coli in only 150 generations.*" (The authors define SP as follows: "*We term pathways constructed from promiscuous activities "serendipitous pathways" (SPs) to denote their fortuitous nature*.")

Authors of Ref. 52 also observed similar high adaptability of biochemical mechanisms: "*These results demonstrate that E. coli can adapt to the loss of a major metabolic gene product with only a handful of mutations and that adaptation can result in multiple, alternative phenotypes*."

"*We also found evidence showing that adaptive evolution can lead to multiple, alternative phenotypes within the evolutionary landscape, defined in this study as a difference in byproduct secretion rates."*

The following quote emphasizes that the biochemical adaption, considered in the quoted papers, affects rather the *whole* biochemical machinery.

"*This result implies that adaptation required global, network-level changes to transcriptional regulation and metabolism, which is emerging as a general, recurring theme across multiple organisms*." If we recall a quote about genetic switches from [2], that would mean a drastic rewriting of such genetic switches, and who is then responsible for that, what a mysterious force acts behind the scene? Certainly these are not genetic switches themselves, which were much damaged and partially destroyed already.

Another interesting observation comes from Ref. 54 for *S. pombe*: "*As in other organisms... genes in the subtelomeric regions showed low essentiality (1.2%) compared to a genome average of 26.1%.*" Such indeed great redundancy (being it 1.2 or 26.2%) forces to think of genes as of some "pool of opportunities", managed at least in part by some external forces, but not biochemical mechanisms alone.

Thus, the presented results show that the biochemical machinery possesses high adaptability and great variability of newly created pathways, which propagate *globally* through the whole organism.

### 10.2. *Rate of growth of recovered (evolved) organisms*

Now the most interesting part comes - do evolved organisms reach the growth rate defined by the growth ratio? We can only answer indirectly to this question, since the experimenters did not measure parameters necessary for calculation of the growth rate by the growth equation. However, what we found already is that (a) the growth rate for the studied organisms (*E. coli* and *S. pombe*) is defined by the growth ratio (otherwise the experimental dependencies and theoretical curves on Figs. 5 and 7 won't coincide). Next (b), the recovered organisms did have the growth rate close or very close to the original one, and the rates of growth for different strains for the same growth conditions *were equal*.

Some difference of the growth rates in the original wild organisms and in evolved ones with initially damaged biochemical machinery is justified, since the evolved strains may have different sizes and variations in a geometrical form. Let us start from Ref 52 (bold font is added by the author).

"***despite similar growth rates****, at least three distinct endpoint phenotypes developed*". To us, the "*similar growth rates*" are of importance.

"*The growth rates ... of 50-day evolved Dpgi mutants increased 3.6-fold on average, respectively, after adaptive evolution*."

"*All ten replicates converged to* ***similar endpoint phenotypes*** *after fifty days of serial passage and adaptive evolution in glucose M9 minimal media* ***when assessed for changes in growth rates.***"

"*On average, the growth rates for the ten strains exhibited a 3.6-fold increase over the starting unevolved Dpgi strain to a final value of 0.50* $hr^{-1}$ *. That growth rates of parallel replicates converge during adaptive evolution has been observed previously and* ***thus seems to be a reproducible phenotypic outcome*** *of such studies* [17]" (Reference [17] in this quote corresponds to our Ref. 55.) Note that the value of 0.50 $hr^{-1}$ approximately corresponds to the rate of growth of wild type species, i. e. to the original species. Thus, in this case, we have not only equal growth rates of different evolved strains, but also rate of growth close to the one defined by the growth ratio. The note "*thus seems to be a reproducible phenotypic outcome of such studies*" is an important one, since it generalizes the obtained results across different studies, which much contributes to our thesis that rate of growth in evolved strains tends to reach the values defined by the growth ratio. (Of course, hypothetically, there could be another mechanism, or a factor, producing a similar invariant, but such a mechanism is unknown and never emerged in so many studies.)

Thus, the presented results, indeed, enforce our idea that the growth rate in wild and evolved unicellular organisms is defined by the growth ratio. In unevolved organisms, the growth ratio, or the relative amount of synthesized biomass, is that factor, which defines the *direction*, in which composition of biochemical reactions of damaged organism changes towards.

Another revealing thing this section showed is how very flexible with regard to even very substantial damage, like removal of genes, the biochemical machinery of considered organisms is. From systemic point of view, such indeed amazing level of flexibility and adaptability is impossible to achieve for such an enormously complex system using rigid rules, implemented by "genetic switches". Self-organization is one answer, but that rather creates more questions, and especially the main one - which cause forces this self-organization of different strains to move towards the same growth rate?

Finally, the last note. One more validation could be this. If we could make such a hypothetical biochemical damage that the rate of biomass synthesis *increases*, then in the evolved species the rate of growth should be *reduced* to the value defined by the growth ratio. However, it will be very difficult to implement such increase, if at all. Maybe cancer cells could provide some insights in this regard.

## 11. A physical-biochemical paradigm of Life as a comprehensive and transparent cognitive framework

Every new development brings new questions. Scientific ideas, by definition, are prohibited to be carved in stone. Given its generality and omnipresence, the discovery of the general growth mechanism brings lots of questions, which have to be answered in order to move forward. The main idea behind such questioning is to have a clear and entirely transparent understanding of fundamental mechanisms defining Life, and a clear vision of fundamental concepts underlying both known and yet undiscovered mechanisms. There should not be blank spots filled with "obvious" assumptions (which in most instances actually turned out to be *beliefs*), similar to an idea about the exclusive managerial role of some genetic code responsible for everything. The updated physical-biochemical paradigm of Life provides directions for such studies and all sorts of opportunities to make them meaningful, efficient and successful.

For that, Life should be understood as an *entirely automated autonomous process and mechanism* at the same time, in which all constituting factors and mechanisms fit together, and each one is fully exposed to the observer in its principles, appearance and action. One should be able to understand *why*, not only to see *how* (as it happens now) this mechanism takes raw materials in certain conditions as an input, and, passing them through the process, delivers a living organism at the output. The analogy could be a flight of a thrown ball: we know that before the flight it was accelerated with certain force acting at a certain distance, and so the launching speed is described by such and such equation of classical mechanics. Then, the ball starts slowing its motion. Yes, that's right, says one - that because of the force of gravity, which acts on it, pulling it approximately to the center of the Earth, which is for this latitude defined by such and such formulas. Then, yes, friction of the air has to be taken into an account, which is defined by such and such equations, whose parameters we can find using such and such measurements. And so on. At every moment of flight, we know what laws of nature act, why and how they work, and what quantitative apparatuses describe them, leaving no gaps for unfounded assumptions.

One may not know some things, but there should be the right conceptual *framework* to acquire such understanding. Does the present biochemical paradigm allow doing so? The answer is 'No', since it apriori excludes possible mechanisms, which might act at higher than molecular levels, while such mechanisms, indeed, could exist, given the known physical arrangement of the world. Thus, there is a definite need for an improved, more comprehensive and more transparent cognitive and methodological paradigm of Life, incorporating all relevant fundamental principles, and this is what the proposed physical-biochemical paradigm of Life provides.

## 12. Conclusion

The article began with a discussion of a fundamental question - What governs Life origin and development? Different, often opposing, views of different levels of generality on possible mechanisms were considered. We followed arguments of E. Schrödinger in his famous lectures and found what led him to make the assumptions, which significantly influenced the following course of this discipline. The truth is that Schrödinger did not exclude an idea that Life is governed by an 'ordinary' or special physical laws. His assumptions, as we have seen, do not contradict our findings - the general growth mechanism indeed is an 'ordinary' physical law, or mechanism. Is it 'special'? To a certain degree, yes, since it is applicable to living matter, although not at molecular, but at higher spatial levels. Another very plausible, and at the same time *very important*, option, which E. Schrödinger did not consider at all, is that Life is governed by *several* physical and biochemical mechanisms simultaneously, acting in cooperation. It is namely *this* arrangement, which we find in Nature so omnipresent.

There is nothing wrong with having doubts and even making erroneous conclusions when solving so complicated fundamental level problems. Search for truth is by definition iterative and incremental. It's just important to continue moving *forward*, whatever the state of affairs is, regardless of how final, perfect and satisfying it might look at a first glance. Motion is an inherent property of matter as such; it never stops moving. No paradigm, no concept can be final, representing the end of story. When such a situation happens, it just means that the paradigm or idea became a *dogma*, which from that moment becomes an obstacle in the following development. It does not mean that such a paradigm has to be necessarily rejected. Normally, the right things become foundations for the following progress. In case of theories or paradigms, such prior developments, whole or in part, are included into the newer more general theories and paradigms, thus securing the succession and continuity of knowledge. (If this does not happen, then the discipline is in trouble, as it happened with physics of elementary particles, or with philosophy, once the classic German philosophy, the jewel of human intelligence and reasoning, was rejected as the basis for the following development, and numerous opportunists began proposing their new "philosophies" from scratch.) In case of the proposed physical-biochemical paradigm of Life, all known biochemical mechanisms are included into it, presenting an inherent, inseparable part of a newer Life paradigm. It's just that the biochemical mechanisms play a somewhat different role in it.

From the wide range of possibilities which Schrödinger considered, only the idea of special properties of living matter was accepted, and then it was amplified and modified to the status of an exclusively biochemical paradigm of Life origin and development. The idea, apparently, appealed to the way of thinking many people adhere to, which is extracting few aspects from a multifactor phenomenon, and stick with them, ignoring the rest of factors and evidences. Despite the substantial opposition among researchers, the unilateral biochemical paradigm of Life eventually prevailed, favored by grants, high ranking publications, prestige awards, etc.

The opponents, accordingly, were suppressed through the same means. It was easier to do so, since they could not present real mechanism, which could support their ideas. Now, the situation is different in this regard: Such a *physical* mechanism, the general growth mechanism, was discovered fifteen years ago, and studied and validated from different perspectives since then. However, the biochemical paradigm is now so solidly established and even canonized in the minds of at least two generations of researchers, raised on this conception (organizationally, educationally, businesswise and in people's mentality, including the general public) that changing the situation will be difficult.

### 12.1. *General growth mechanism. Results, possible applications and validation*

The proposed addition of discovered physical mechanisms to the present biological paradigm was comprehensively verified by different means. *First line* of proofs dealt with the discovery of concrete growth and division mechanisms for particular unicellular organisms. Their validity, or at least very high probability of validity, was convincingly proved by comparison of experimental growth dependencies with the growth curves and characteristics, obtained theoretically on the basis of the growth equation. Similarly, the results of previous research in Refs. 23,24 proved validity of the general growth mechanism, and of the growth equation, for organs.

*The second line* of arguments appealed to high level scientific methodological considerations, especially to criteria for validation of scientific truths. One of the most powerful criteria of validity of general theories is that they should provide convergence to prior obtained more particular results and less general theories. Indeed, we obtained a solution of the growth equation for two particular growth scenarios, which includes a known logistic formula as a particular case of the growth equation Eq. (1) when it was applied to two geometric forms - a disk and a sphere. According to the theory of verification of scientific knowledge, this is a very strong evidence of validity of the general growth mechanism, when previous particular results are included into a new general solution,

Furthermore, we found that the growth ratio for disk and sphere like organisms is an *exponential* function of time, which is also the result important from the validation perspective - many natural processes, indeed, are described by exponential functions, like the attenuation of waves in absorbing media, transitional electric processes, change of atmospheric pressure with height, etc. In this, of especial importance is the fact that the input data - the relative volume and the relative surface of a disk and a sphere - contain no exponential functions, nor even a slight hint to them. The fact that this result is not an abstract mathematical interplay, but the property of *real* physical and biological processes, the property of a *real world*, makes the results even more valuable. Their fundamental meaning is yet to be understood, but it is already clear that some important feature of the real world was discovered, apparently related to its *fundamental* geometrical properties and their relationship with natural exponential processes.

Another important validation consideration is that the present biochemical paradigm with all its experimental results and considerations is *entirely* included into a newer proposed theory – a physical-biochemical paradigm of Life origin and development.

*The third line* of arguments dealt with predictions of certain effects and relationships made on the basis of the general growth mechanism and the growth equation, and their experimental confirmations. Such were the studies of metabolic rates, characteristic behavior of metabolic allometric exponents, growth acceleration and suppression through change of geometry, explanations why unicellular organisms have certain geometrical forms, etc.

*The fourth line* of arguments related to equal growth rate of different strains of recovered (evolved) organisms, whose biochemical machinery was previously damaged. Such equality of growth rate cannot be secured by biochemical machinery alone, while the general growth mechanism very well explains this effect.

The results obtained through these four lines of validation showed validity, or high probability of validity, of obtained results, and consequently of the general growth mechanism and the growth equation.

It is important to emphasize one more time that the proposed physical addition to a biochemical paradigm of Life by no means rejects, but seamlessly incorporates *all*, without exception, biochemical mechanisms discovered within the biochemical paradigm. In this regard, one should not consider the proposed paradigm as an alternative. In fact, this new physical-biochemical paradigm of Life is a *more general*, of the *next qualitative* level, concept of Life origin and development. It accommodates and reconciles all previous studies, while opening new horizons and providing *conceptual* and *methodological* frameworks for the following studies, as well as new perspectives and guidance.

The core of this physical-biochemical paradigm is a general growth mechanism, which is a fundamental physical mechanism. In the same way as laws of classical mechanics are valid in the entire Universe, the general growth mechanism is also an inseparable attribute of the world. It works at a cellular scale level and above, up to the whole organisms, in tight cooperation with biochemical mechanisms, by imposing uniquely defined constraints on the distribution of nutrients between the maintenance needs and biomass production (as well as performing some other tasks we did not discuss here). It universally works in the entire Universe, where conditions allow living organisms to originate.

An additional note about using this *real* physical mechanism and its mathematical representation, the growth equation, for the study of metabolic characteristics of unicellular organisms. We found that (a) obtained values of metabolic allometric exponents, indeed, are located on a straight line in logarithmic coordinates, as it should be if the general growth mechanism and the growth equation are valid; (b) the value of the found allometric exponent complies with experimental observations. Besides, we discovered that the nutrient influx per unit surface also scales as a straight line in logarithmic coordinates for sedentary microorganisms. These results, on one hand, one more time confirm the validity of the general growth mechanism and its mathematical representation, the growth equation. On the other hand, they effectively demonstrate the role of the general growth mechanism as an efficient scientific tool, which allows solving difficult problems and explaining known but still unresolved Nature's puzzles.

Another problem, the general growth mechanism certainly will be very useful for, and actually without which the problem cannot be completely solved *in principal*, is a much debated topic of safety and consequences of consuming genetically modified food by people and domesticated animals. We won't discuss the issue here, but even at the present stage the general growth mechanism can provide principal insights for this problem, indicate the optimal directions and methods for its study and - to some extent - foresee the results of such studies and possibly outline the potential issues.

Next area of application, for which the proposed physical-biochemical paradigm of Life would be extremely useful, is evaluating probabilities of extra terrestrial Life in different environments, on different time scales, possible development scenarios, etc, not mentioning the answer to the main question, is extra terrestrial Life possible? - which is still unknown. The general growth mechanism already much contributes towards definitive answer 'yes'.

### 12.2. *Importance of the general growth mechanism and growth equation for biological studies and following discipline's progress*

It might seem puzzling finding so many important characteristics, explanations and applications from a single growth equation introduced heuristically. However, this is how fundamental mechanisms in nature *always* work. What is needed to compute the trajectories of a thrown stone or a planet? By and large, only the Newton's Second law of mechanics, represented by a simple equation with three linear terms, $a = F/m$ (acceleration, force and mass). The general growth mechanism is also a mechanism from the same category of generality, so that such "fertility" should not come as a surprise, but actually *has to be* expected. The only difference is that the general growth mechanism is more complicated and more difficult for understanding (otherwise, it probably would be discovered already). Such a generality and wide range of applications of the general growth mechanism is also an additional proof of its validity. If it was not a real mechanism, it would fail at a first application attempt, say when modeling experimental growth dependencies for *Amoeba*, and even more so at an attempt finding its division mechanism. However, in fact, we observed the opposite, when, following through a *whole* long chain of events associated with division, we have seen that at every division phase the general growth mechanism and the growth equation very well explain the observed effects and known facts about *Amoeba* division, and not only *qualitatively*, but also *quantitatively*, and with a high accuracy. The same successful outcome was observed for growth and division models of other (very different, by the way, such as *S. pombe*) organisms through all their growth and division phases as well.

One more argument in favor of validity of the general growth mechanism and other considered physical mechanisms is that there are many biological phenomena, which originate at a higher than biomolecular level. Ref. 33 gives an example of how the cell size matters for the metabolic properties of multicellular organisms - the result, which cannot be derived from biochemical mechanisms. The results on livers and liver transplants' growth in dogs and humans, obtained using the general growth mechanism and the growth equation in Refs. 23,24, *principally* cannot be obtained on a biochemical basis.

The general growth mechanism, in fact, makes understanding of life origin and development much simpler; it removes the aura of mystery surrounding the notion of preprogrammed genetic codes, thus reducing the entire phenomena to transparent cooperative workings of physical and chemical mechanisms. On phenomena of such a scale as Life, Nature must work on simpler, more elegant, and much more reliable and optimal principles than relying on particular molecular events; and this cannot be otherwise, as all previous knowledge, especially the one coming from physical laws and natural philosophy, witnesses. By and large, biochemical mechanisms are necessary *executors* - very active, persistent, sending feedbacks, with lots of possibilities and workarounds; tireless and absolutely indispensable foundation of life. But not the only one.

At some point, the presently fragmented biochemical mechanisms have to be united on a more general basis (or bases). The general growth mechanism and the proposed physical-biochemical paradigm of Life, presented here, include and unify these mechanisms, as well as all other known facts and knowledge, at *all* scale levels. In short, the main features of this physical-biochemical paradigm can be described as follows:

*Life is governed by cooperative working of physical and biochemical mechanisms. The discovered general growth mechanism, acting at cellular level and above, imposes macro-constraints, defined by geometry of organisms, and mathematically expressed through parameters of the growth equation. The value of one of these parameters, called the growth ratio, is directly tied to geometrical form, and defines the relative amount of produced biomass (relative to the nutrient influx used for maintenance). Biochemical mechanisms adjust to imposed macro-constraints at each moment of growth and reproduction. The major constrain, the relative amount of produced biomass, defines composition of biochemical reactions, while the biomass increase, in a feedback manner, causes the change of organism's geometry, which changes the growth ratio, which in turn changes ratio of fractions of nutrients used for biomass synthesis and maintenance. The change in relative amount of produced biomass, besides altering the composition of biochemical reactions, also secures their irreversibility. Through this continuous feedback loop organisms are forced to autonomously, automatically, and generally irreversibly proceed through successive phases of their growth and reproduction cycles.*

People familiar with physics and natural philosophy, will immediately recognize the pattern common to *all* fundamental laws and mechanisms of Nature. What all these laws and mechanisms, by and large, do? From all possible numerous scenarios, paths and opportunities, they define an *optimal*, *uniquely* defined path (based on criteria, specific for an area of application of the law), thus introducing *stability* and *order* into this world. Exactly the same thing does the general growth mechanism, when it introduces a uniquely defined *invariant*, the growth ratio, which from all possible distributions of nutrients between the biomass synthesis and maintenance defines *the only* optimal path, and thus also creating stability and order in the world of living organisms. (In the same way, Hamiltonian through generalized coordinates in classical mechanics defines *the only* possible trajectories of systems of moving material bodies.) Imagine what would happen with a multicellular organism if it was not the case; imagine what would happen with populations of species having arbitrary periods of reproduction and absolutely no notion of a *generation*. Fortunately, this is not the case, and there is much more stability and order in the real living world, and this is why it existed for so long, subjected to objective *laws of Nature*, overcoming (with a growing difficulty though, as the power of humankind increases) the subjectivity of humans, which is not rarely incapable to adequately enough reflect on the laws and mechanisms of Nature, governing the objective reality, and so providing erroneous feedbacks-actions.

## Acknowledgments

The author warmly thanks Professor P. Pawlowski for the many years support of this study; Dr. A. Y. Shestopaloff for the help with prior editing and productive discussions and Dr. K. Y. Shestopaloff for the help with the literature.